\documentclass[10pt, a4paper]{article}

\usepackage{graphicx}
\usepackage[dvipsnames, svgnames, x11names]{xcolor}
\usepackage[margin=1in, bottom=1in]{geometry} 
\usepackage[hidelinks]{hyperref}
\usepackage{fancyhdr} 
\usepackage{authblk} 
\usepackage{tcolorbox} 
\usepackage[super,sort&compress]{natbib} 
\setcitestyle{citesep={,}}
\usepackage{float} 
\usepackage[nolist]{acronym} 
\usepackage{multirow}
\usepackage{mathtools, nccmath} 
\usepackage{booktabs} 
\usepackage{caption}
\usepackage{amssymb} 
\usepackage{rotating} 
\usepackage{siunitx}
\usepackage{ragged2e} 
\usepackage{wrapfig}
\usepackage{makecell}

\definecolor{SBGold}{HTML}{EBB028}
\definecolor{SBLightGrey}{HTML}{D9D9D9}
\definecolor{SBDarkGrey}{HTML}{B3B3B3}

\begin{acronym}
    \acro{ID}{in-domain}
    \acro{OOD}{out-of-domain}
    \acro{OC20 dataset}{Open Catalyst 2020 dataset}
    \acro{MD}{molecular dynamics}
    \acro{TS}{transition state}
    \acro{DFT}{density functional theory}
    \acro{VASP}{Vienna Ab Initio Simulation Package}
    \acro{RPBE}{revised Perdew-Burke-Ernzerhof}
    \acro{MP}{Materials Project}
    \acro{MLIPs}{Machine learning interatomic potentials}
    \acro{MAE}{mean absolute error}
    \acro{MLP}{multilayer perceptron}
    
\end{acronym}

\newcommand{\CompanyLogo}{\includegraphics[width=0.25\textwidth]{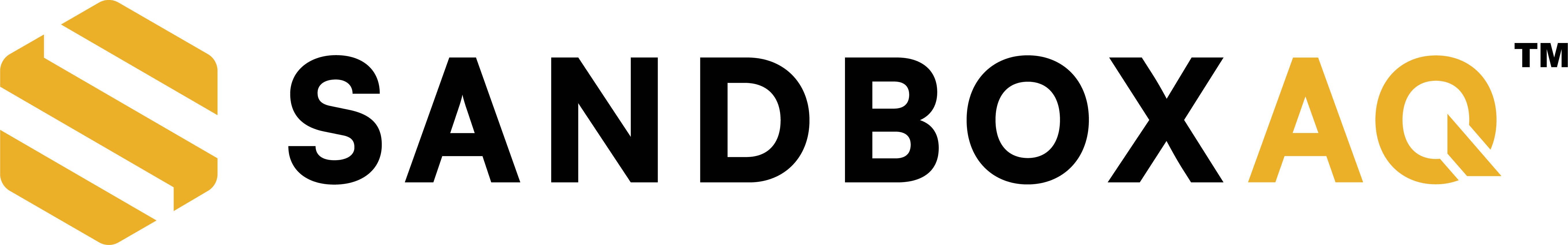}}
\newcommand{\CompanyLogoSmall}{\includegraphics[width=0.15\textwidth]{SandboxAQ-Logo-Accent-Color-Black-TM.png}}

\newcommand{\sandboxaq}{SandboxAQ Core Team, Palo Alto, CA}
\newcommand{\sandboxaqsupport}{SandboxAQ Support Team, Palo Alto, CA}
\newcommand{\nvidia}{Nvidia Corporation, Santa Clara, CA}

\newcommand{\coauthorsymbol}{\textdagger} 
\newcommand{\correspondingsymbol}{*}

\DeclarePairedDelimiter{\nint}\lfloor\rceil

\newcommand{\PaperTitle}{AQCat25: Unlocking spin-aware, high-fidelity machine learning potentials for heterogeneous catalysis}

\newcommand{\customauthors}{%
    \begin{center}

        Omar Allam\textsuperscript{1,\coauthorsymbol},
        Brook Wander\textsuperscript{1,\coauthorsymbol},
        SungYeon Kim\textsuperscript{1},
        Rudi Plesch\textsuperscript{2},
        Tyler Sours\textsuperscript{2},
        Jia-Min Chu\textsuperscript{2},
        Thomas Ludwig\textsuperscript{2},
        Jiyoon Kim\textsuperscript{2},
        Rodrigo Wang\textsuperscript{2},
        Shivang Agarwal\textsuperscript{2},
        Alan Rask\textsuperscript{2},
        Alexandre Fleury\textsuperscript{2},
        Chuhong Wang\textsuperscript{2},
        Andrew Wildman\textsuperscript{2},
        Thomas Mustard\textsuperscript{2},
        Kevin Ryczko\textsuperscript{2},
        Paul Abruzzo\textsuperscript{3},
        AJ Nish\textsuperscript{3}, \\
        Aayush R. Singh\textsuperscript{1,\correspondingsymbol}\\
        \vspace{1em}

        \small \textsuperscript{1}\sandboxaq \\
        \vspace{0.5em}
        \small \textsuperscript{2}\sandboxaqsupport \\
        \vspace{0.5em}
        \small \textsuperscript{3}\nvidia \\
        \vspace{0.5em}
    
        \small Correspondence: aayush.singh@sandboxquantum.com
     \end{center}
 }

\newcommand{\AbstractText}{Large-scale datasets have enabled highly accurate machine learning interatomic potentials (MLIPs) for general-purpose heterogeneous catalysis modeling. There are, however, some limitations in what can be treated with these potentials because of gaps in the underlying training data. To extend these capabilities, we introduce AQCat25, a complementary dataset of 13.5 million density functional theory (DFT) single point calculations designed to improve the treatment of systems where spin polarization and/or higher fidelity are critical. We also investigate methodologies for integrating new datasets, such as AQCat25, with the broader Open Catalyst 2020 (OC20) dataset to create spin-aware models without sacrificing generalizability. We find that directly tuning a general model on AQCat25 leads to catastrophic forgetting of the original dataset's knowledge. Conversely, joint training strategies prove effective for improving accuracy on the new data without sacrificing general performance. This joint approach introduces a challenge, as the model must learn from a dataset containing both mixed-fidelity calculations and mixed-physics (spin-polarized vs. unpolarized). We show that explicitly conditioning the model on this system-specific metadata, for example by using Feature-wise Linear Modulation (FiLM), successfully addresses this challenge and further enhances model accuracy. Ultimately, our work establishes an effective protocol for bridging DFT fidelity domains to advance the predictive power of foundational models in catalysis.}

\newcommand{\makeinfobox}{%
    \begin{tcolorbox}[
        colback=SBLightGrey, 
        colframe=SBGold, 
        boxrule=1pt, 
        arc=5mm, 
        left=10mm, right=10mm, top=10mm, bottom=10mm, 
        boxsep=0pt, 
        width=\textwidth 
    ]
        
        \centerline{\parbox{0.95\linewidth}{ 
            \centering 
            \LARGE\bfseries\PaperTitle 
        }}
        \vspace{1.5em} 
        
        {\large\customauthors\par} 
        \vspace{1.5em}

        \vspace{-1ex} 
        \normalsize 
        \justify 
        \noindent\ignorespaces 
        \AbstractText 
        \par 
        \vspace{1.5em} 
        
        \centerline{\CompanyLogo}
    \end{tcolorbox}
    \vspace{1em} 
}

\fancypagestyle{plain}{%
    \fancyhf{} 
    \fancyhead[L]{\CompanyLogoSmall} 
    \fancyhead[R]{}

    \fancyfoot[L]{\footnotesize\textsuperscript{\coauthorsymbol}Co-first authors; \textsuperscript{\correspondingsymbol}Corresponding author}
    
    \fancyfoot[C]{}
    \fancyfoot[R]{}
}

\date{\today}

\begin{document}
\thispagestyle{plain}

\makeinfobox

\section*{Introduction}

Over the past three decades, computational approaches that couple first-principles \ac{DFT} with microkinetic modeling have become a cornerstone of modern heterogeneous catalysis research by providing a framework for rational catalyst design \cite{norskov2009towards, greeley2016theoretical, motagamwala2020microkinetic, chen2020computational, xie2022achieving}. Numerous studies have linked atomic-scale surface chemistry to macroscopic kinetic observables, enabling the elucidation of complex reaction mechanisms for a variety of critical industrial heterogeneous catalytic processes including, but not limited to, ammonia synthesis \cite{honkala2005ammonia, singh2018computational}, methanol synthesis \cite{grabow2011mechanism, studt2012co}, Fischer-Tropsch synthesis \cite{van2013mechanism, yao2019quantitative}, selective hydrogenation \cite{studt2008identification, he2021catalyst}, steam reforming of methane \cite{jones2008first, zhu2021first}, the water-gas shift reaction \cite{schumacher2005trends, ghanekar2020catalysis}, and ethylene epoxidation \cite{stegelmann2004microkinetic, li2021understanding}. Many of these studies have leveraged unifying concepts such as d-band theory and the Br{\o}nsted-Evans-Polanyi relations \cite{hammer2000theoretical, norskov2002universality, bligaard2004bronsted, van2009reactivity}, which correlate the binding and transition-state energies of elementary reactions, facilitating the construction of volcano plots that predict optimal catalyst performance, in some cases even leading to experimentally validated discovery of new catalysts \cite{besenbacher1998design, jacobsen2001catalyst}. Despite these successes, the prohibitive cost of \ac{DFT} largely limits its application to relatively simple networks of reactions taking place over idealized low-index facets of unary and binary catalyst materials \cite{greeley2016theoretical, van2009reactivity}.

\begin{figure}[ht]
    \centering
    \includegraphics[width=\linewidth]{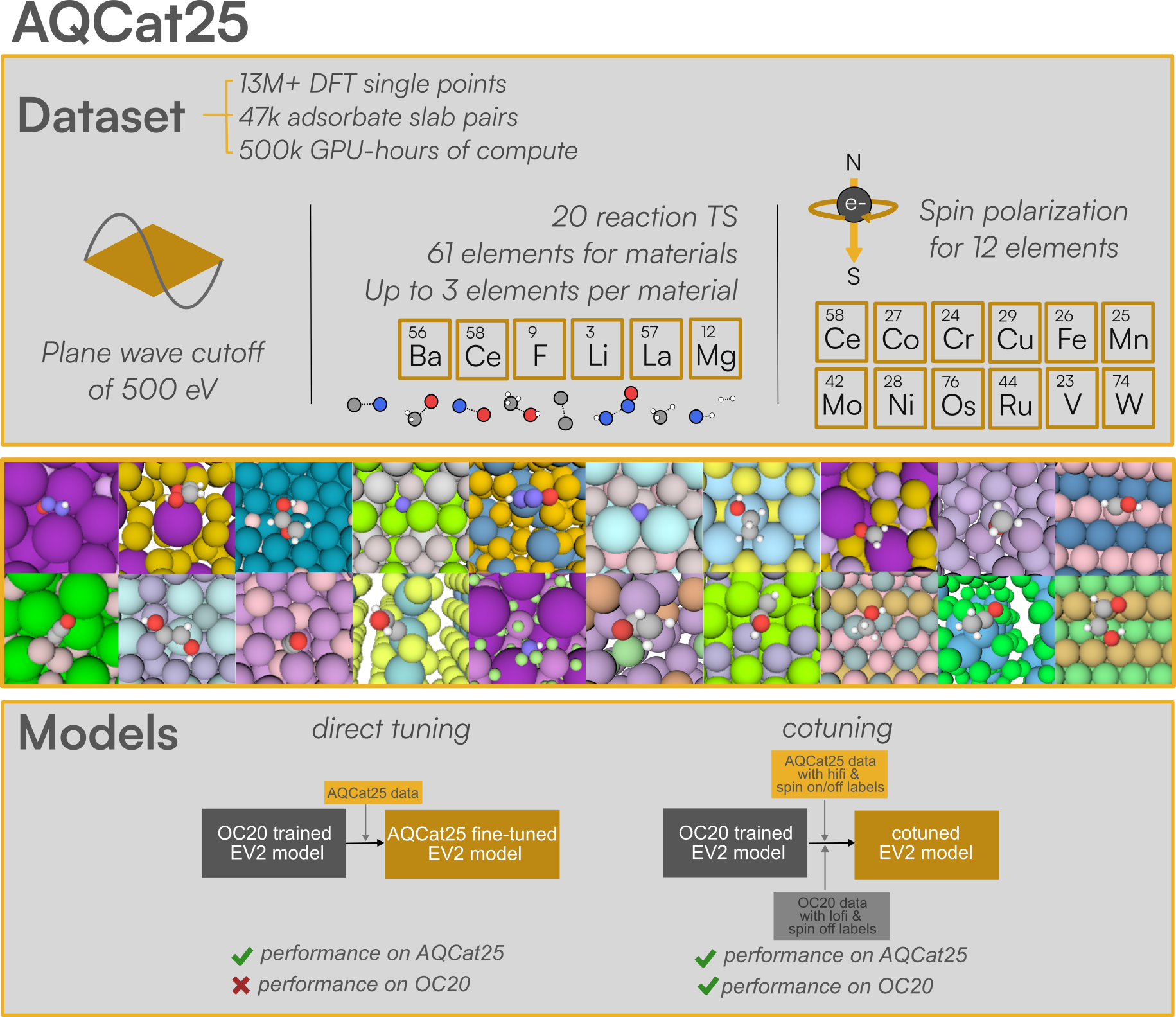}
    \caption{A summary of the AQCat25 dataset and models. Spin polarization has been included for 12 important elements. A plane wave cutoff of 500 eV is used. Six new elements are added, when compared to the \ac{OC20 dataset}\cite{chanussot2021open} as well as 20 transition state adsorbates. Models were jointly tuned/trained from scratch on OC20 and AQCat25 to achieve good performance on their validation and test splits, respectively.}
    \label{fig:summary_fig}
\end{figure}

\ac{MLIPs} have emerged as an attractive alternative to estimate electronic structure properties at near-quantum accuracy for a small fraction of the computational cost \cite{deringer2019machine, unke2021machine, pablo2023fast,tang2024machine, bozal2025developing, yuan2025foundation}. These models learn the interactions required to predict the potential energy landscape of atomistic systems from large-scale public databases of \ac{DFT} calculations \cite{chen2021learning, merchant2023scaling, batatia2023foundation, ko2025data}. State-of-the-art models for heterogeneous catalysis are made possible by Meta FAIR's Open Catalyst 20, 22, and 25 datasets \cite{chanussot2021open, tran2023open, sahoo2025open}, which collectively consist of nearly 300 million single-point DFT calculations of adsorbate-surface interactions relevant for reactions of carbon, hydrogen, oxygen, and nitrogen over a diverse catalyst space spanning most of the periodic table. The introduction of machine learning methods into computational catalysis workflows has begun to enable studies of reaction network complexity \cite{ulissi2017address, zhong2020accelerated, margraf2023exploring, morandi2024foundational} and catalyst structural dynamics \cite{mou2023bridging, owen2024atomistic, omranpour2025machine} that were completely inaccessible just 10 years ago.

Although the sheer scale of these datasets has necessitated some compromises in the fidelity of the underlying training data, convergence of the resulting adsorption energies benefits from error cancellation and has been validated with respect to most \ac{DFT} settings, with plane-wave cutoff and smearing width requiring some improvements to accurately capture total energies of non-metals \cite{abdelmaqsoud2024investigating}. One of the most significant gaps in existing large-scale heterogeneous catalysis datasets is the treatment of magnetism. Since spin-polarized \ac{DFT} calculations are considerably more expensive than spin-unpolarized calculations, spin is often omitted in the interest of scale and throughput \cite{xu2025spin}. The consequence of this choice is that the resulting models are not suitable for many industrially relevant catalytic processes such as ammonia synthesis \cite{honkala2005ammonia} and Fischer-Tropsch synthesis \cite{van2013mechanism}, which rely on earth-abundant first-row transition metals (e.g., iron, cobalt, and nickel) that exhibit especially strong spin polarization effects on binding energies and activation barriers \cite{sun2007spin, sun2007spinbenz, cao2023spin, zhang2024spin}. As the field moves towards discovering new, low-cost, and sustainable catalytic materials to replace precious metals, the importance of treating magnetic effects when training foundational \ac{MLIPs} becomes increasingly paramount.

Alongside progress in data generation, developments in machine learning architectures, often based on equivariant graph neural networks, have improved performance on atomistic tasks. For heterogeneous catalysis, models like eSEN\cite{fu2025learning}, EquiformerV2 \cite{liao2023equiformerv2} and EScAIP \cite{qu2024importance} have achieved state-of-the-art results. A significant leap towards broader universality is the Universal Model for Atoms (UMA) \cite{wood2025family}, trained on $\sim$500 million structures across diverse chemical domains (molecules, materials, catalysts). UMA modifies the eSEN architecture with additive embeddings for global context (charge, spin, DFT task) and a Mixture of Linear Experts (MoLE) routed by this context plus element composition. UMA's core design goal is to accurately reproduce the original physics of each training task (e.g., spin-unpolarized OC20), operating as a multi-task surrogate rather than a model explicitly designed to perform cross-fidelity corrections between different levels of theory. A distinct advantage of total energy models like UMA is their ability to better capture, among other effects, restructuring of bare catalyst slabs \cite{abdelmaqsoud2024investigating}.

Other approaches have focused on integrating low- and high-fidelity data for related tasks, such as by augmenting node features with a fidelity one-hot encoding and applying both common and fidelity-specific weights in modified linear layers \cite{kim2024dataefficient} or utilizing a model's intrinsic global state feature to embed fidelity context during message passing \cite{ko2025data}. For example, Ko and Ong demonstrated that a single multi-fidelity model trained with a small fraction of high-fidelity data could achieve similar accuracy to a single-fidelity model requiring eight times the amount of costly high-fidelity training data \cite{ko2025data}. Alternatively, other methods utilize architectural separation, such as dynamically using separate prediction heads branching from a shared backbone for each fidelity level \cite{messerly2025multifidelity, gardner2025understanding}.

While recent work has produced model architectures that can incorporate spin \cite{xu2025spin,chapman2022machine, deng2023chgnet, anstine2023machine, hu2023spin, yang2024deep, yu2024spin}, their predictive power is limited by the absence of high-quality, spin-polarized training data for heterogeneous catalysis. Given the success of methods for adapting models to new data \cite{tran2023open, musielewicz2022finetuna, deng2025systematic, liu2025fine, king2024transfer, radova2025fine}, alongside advancements in training universal models from diverse datasets, we see a clear opportunity to develop improved foundational models specifically for spin-polarized, high-fidelity catalytic systems.

Here, we present the AQCat25 dataset and baseline AQCat25-EV2 models (Figure \ref{fig:summary_fig}), which improve upon the performance of EquiformerV2-31M and EquiformerV2-153M adsorption energy \ac{MLIPs} for heterogeneous catalysis in three key ways: increasing the fidelity of the reference \ac{DFT} calculations, explicitly incorporating spin polarization for magnetic elements, and introducing new elements to the model domain that are underrepresented in existing datasets. Through this work, we demonstrate data-efficient methodologies for building multi-fidelity \ac{MLIPs} that span distinct physical regimes (such as spin polarization) to new domains of chemistry while ensuring that the models maintain accuracy and generalizability across a wide range of catalysts and reactions. The dataset, models, and code are available publicly to support further developments by the academic community.

\section*{Methods}

\subsection*{Density functional theory}
\ac{DFT} calculations were performed using the \ac{VASP}\cite{kresse1994ab,kresse1996efficiency,kresse1996efficient,kresse1999ultrasoft}. A plane-wave cutoff energy of 500 eV was applied, and Gaussian smearing with a width of 0.1 eV was used. The \ac{RPBE}\cite{perdew1996generalized,zhang1998comment} functional was chosen for its performance on heterogeneous catalyst systems, and the system's geometry was optimized using the conjugate gradient algorithm. For systems containing Ce, Co, Cr, Cu, Fe, Mn, Mo, Ni, Os, Ru, V, or W, spin polarization was enabled to account for magnetic effects. For a full list of \ac{VASP} parameters, please see the Supplementary Information.

Although these settings represent a significant increase in fidelity over previous large-scale datasets and are considered nominal for catalysis research, we acknowledge that even higher-fidelity calculations are possible. However, any increase in per-calculation fidelity must be weighed against the loss of dataset diversity for a fixed computational budget. For foundational \ac{MLIPs} that must generalize across a vast chemical space, this trade-off is critical.

\subsection*{Bulk selection}
The AQCat25 bulk materials database was constructed by first updating and then expanding the OC20 dataset. Initially, the \ac{MP}\cite{jain2013commentary,horton2025accelerated} database was queried for all structures containing only elements present in the \ac{OC20 dataset}, subject to the constraints of a maximum of three unique elements per material and an energy above the convex hull ($E_{\text{hull}}$) of 0.1~eV/atom or less. Structures from the original \ac{OC20 dataset} not found in this query were retained only if they were structurally unique. To expand the chemical space, a second query was performed using the same stability and size constraints but including six additional elements: Li, Ba, La, Ce, Mg, and F. The resulting set of new materials was then subsampled to ensure balanced representation. Up to 500 structures were randomly selected for each group containing a single new element, and up to 20 structures for each group containing a combination of new elements. The dataset was assembled by combining the updated \ac{OC20 dataset} materials, the preserved unique structures, and the sampled new materials. Finally, the dataset was filtered to only contain bulks with up to 30 atoms per unit cell. Data splits were then assigned by attempting to preserve the original designations for all \ac{OC20 dataset} materials and distributing new materials based on chemical composition to maintain consistency with the established \ac{OC20 dataset} splitting methodology.

\subsection*{Adsorbate-slab selection}
The number of single points and systems that make up the splits and data types included in the AQCat25 dataset is shown in Table \ref{tab:dataset_size_table}. Here, a system is defined as a unique adsorbate-slab pair for that subsplit.

\subsubsection*{Dataset splits}
\begin{table}[H]
    \centering
    \footnotesize
    \begin{tabular}{c|l|c|c}
        Primary split & Secondary Split & N systems & N single points \\ \hline
                    &  Relaxations & 24,624 & 6,959 k \\ 
         In Domain  & Rattled & 8,189 &  947 k \\
                    & Transition states & 2,854 & 676 k \\
                    & Molecular Dynamics & 2,098 & 249 k \\ 
                    & OC20 fidelity, spin on relaxations & 4,831 & 863 k   \\ \hline
                    & OOD adsorbate relaxations & 1,913 & 577 k \\
         Validation & OOD material relaxations & 991 & 318 k \\
                    & OOD both relaxations & 994 & 295 k \\ \hline
                    & OOD adsorbate relaxations & 992 & 347 k \\
         Test       & OOD material relaxations & 994 & 316 k \\
                    & OOD both relaxations & 988 & 356 k \\ \hline
                    & ID & 19,273 & 1,282 k \\
                    & ID OC20 fidelity, spin on & 4,868 & 273 k \\
         Slabs      & OOD validation & 497 & 29 k \\
                    & OOD test & 498 & 36 k \\ \hline
         
         \multicolumn{2}{c|}{Totals} & 47 k & 13.5 M \\

    \end{tabular}
    \caption{The number of systems and single points across data splits. The total system count reflects the number of unique adsorbate-slab combinations.}
    \label{tab:dataset_size_table}
\end{table}
The dataset is structured into three primary splits: \ac{ID}, \ac{OOD} validation, and \ac{OOD} test. Each \ac{OOD} split is further categorized by the type of novelty introduced, either in the adsorbate or in the material slab. This strategy is designed to evaluate the model's ability to generalize to novel systems it has not seen during training, in the same manner as the \ac{OC20 dataset}\cite{chanussot2021open}. The \ac{ID} split contains configurations where both the adsorbate and the material slab are present in the training set. The test and validation \ac{ID} splits serve as a baseline for the model's performance on familiar data and are sampled from the same distribution of the training set. Both \ac{OOD} splits are designed to test the ability of machine learning models to generalize. The \ac{OOD} validation set is used for hyperparameter tuning, while the \ac{OOD} test set provides a final, unbiased evaluation of the model's performance on unseen data.

The following categories are included in both \ac{OOD} splits: (1) \ac{OOD} adsorbate, (2) \ac{OOD} material, (3) \ac{OOD} both. For \ac{OOD} adsorbate, the material slab is \ac{ID}, but the adsorbate is new and does not appear in the training data. The test \ac{OOD} adsorbates also do not appear in the validation split and the validation \ac{OOD} adsorbates also do not appear in the test split. For \ac{OOD} material, the adsorbate is \ac{ID}, but the bulk lattice structure (not necessarily its composition) used to construct the slab is new and does not appear in the training set. For \ac{OOD} both, the adsorbate and the material slab are new and do not appear in the training set. The same segregation for validation and test also applies.

\subsubsection*{Sampling diverse states}

To ensure models trained with this dataset have a robust understanding of different structural and energetic states, we employed several calculation types for data generation. The dataset samples both high-energy, off-equilibrium states and low-energy, near-equilibrium states. To sample low-energy states we performed adsorbate-slab structure relaxations. Relaxation calculations involve iteratively optimizing atomic positions to find a local energy minimum. A \ac{DFT} call is made to determine forces, and atoms are moved along these force vectors. This process is repeated until the maximum force on any atom is less than 0.03 eV/$ \text{\AA} $~ or a maximum of 800 steps are reached. These trajectories sample a range of configurations from high to low forces. All \ac{OOD} validation and test set calculations are relaxations. 

To sample high-energy states we took three approaches: (1) running \ac{MD} calculations, (2) placing \ac{TS} systems, and (3) rattling atoms. To sample high-energy states accessible at elevated temperatures, we performed \ac{MD} calculations. Starting from a relaxed structure, we ran 80 steps of \ac{MD} at 900 K. To provide the model with examples of highly distorted configurations relevant to chemical reactions, we extracted transition state structures from the OC20NEB dataset\cite{wander2025cattsunami}. These adsorbates were placed on new surfaces, followed by a short 5-step relaxation. This process generates data with high forces and energies, supporting the training of models that can handle reactive states. To further augment high-force data, we generated rattled configurations by randomly perturbing atomic positions. Two methods were used: (1) rattling all atoms and (2) rattling only adsorbate atoms, with displacements sampled from a normal distribution ($\sigma$ = 0.05, 0.1, 0.15, or 0.2 $ \text{\AA} $). Some rattled systems underwent a single \ac{DFT} calculation, while others had a short 5-step relaxation. Systems whose max absolute force or absolute adsorption energy exceeds 50 eV/Å and 10 eV were excluded from training and evaluation.

\subsubsection*{Additional data}
To explore the opportunity to train models with less costly \ac{DFT} data, we considered data that include spin polarization but with settings that otherwise match the \ac{OC20 dataset}\cite{chanussot2021open}. Notable differences between this data and the rest of the AQCat25 dataset are that we used a plane wave cutoff of 350 eV and Methfessel-Paxton smearing with a width of 0.2 eV. This data aids in understanding how the model handles the distinct physical regimes defined by fidelity and spin polarization. This dataset complements the existing high-fidelity spin-on/off (AQCat25) and spin-off OC20 data by filling a missing quadrant. Adsorption energies were computed using high-fidelity adsorbate references for all spin on systems. We found this to have little impact on the final target energies from preliminary tests.

We also wanted to form an understanding of model performance on the task of finding the minimum adsorption energy for an adsorbate-slab combination. To do this we constructed a small dense dataset, similar to the OC20dense dataset presented by Lan et al.\cite{lan2023adsorbml}. For this dataset we selected 109 adsorbate-slab pairs. Adsorbates were selected to be disassociation reactants from the OC20NEB dataset\cite{wander2025cattsunami}. Slabs were selected randomly, but we selected the materials they were cut from more strategically. We included five unary materials, five binary non-metal materials, 46 binary intermetallics, 30 ternary intermetallics, and 23 ternary non-metals. Within these categories, the bulks were also randomly selected from the bulk database. For each adsorbate-slab pair, we performed 50 placements using the random site with heuristic placement mode in \texttt{fairchem} \cite{fairchem}. These placements were relaxed with the same \ac{DFT} settings as the broader AQCat25 dataset. The relaxed states were filtered using the same algorithms presented by Lan et al.\cite{lan2023adsorbml} to find desorption, dissociation, intercalation, and significant surface change.

\subsubsection*{System enumeration}
All systems were prepared using the publicly available \texttt{fairchem} package \cite{fairchem}. Slab enumeration was performed using the underlying \texttt{pymatgen}\cite{sun2013efficient, tran2016surface} algorithm. Adsorbate placement was performed heuristically at random sites. Rattled systems were perturbed after adsorbate placement using the rattle functionality in \texttt{ASE}\cite{ase_old, ase-paper}. For \ac{TS} systems, they were placed as normal adsorbed intermediates would be by preparing a new adsorbate database with \ac{TS} entries.

\section*{Machine Learning Experiments}

A challenge in this work is training a single MLIP that can accurately predict energies and forces across a dataset containing multiple \ac{DFT} settings. The combined training data spans four distinct physical regimes: high-fidelity spin-on, high-fidelity spin-off, low-fidelity spin-on, and the original low-fidelity spin-off. We therefore explored methods to introduce this \ac{DFT} context, namely spin treatment and calculation fidelity, directly to the model. Inspired by the effectiveness of Feature-wise Linear Modulation (FiLM) \cite{perez2018film} and similar successful techniques \cite{wood2025family} for multi-task learning, the baseline models presented in this paper focus on adapting the EquiformerV2 (EV2) architecture\cite{liao2023equiformerv2} using this approach. We did not focus on fine-tuning UMA\cite{wood2025family} models because of licensing issues, but we expect those to have even better performance than the models presented here. The Feature-wise Linear Modulation (FiLM) technique \cite{perez2018film} provides an expressive conditioning mechanism. Rather than adding context via feature concatenation, FiLM applies a learned, feature-wise affine transformation ($\gamma F + \beta$) that can scale, shift, or suppress activations. This strategy of deep, additive modulation is similar in principle to the additive embedding mechanism successfully employed by the UMA family of models \cite{wood2025family}. To evaluate performance on AQCat25 while retaining knowledge from OC20, we used the EV2\cite{liao2023equiformerv2} model architecture with three variants and three training protocols. The variants were: (i) \textbf{EV2} (unmodified), (ii) \textbf{EV2-inFiLM}, which applies additive FiLM \cite{perez2018film} shifts to the scalar ($l{=}0$) channels at the input, and (iii) \textbf{EV2-in+midFiLM}, which applies the same modulation at the input and after each equivariant block. The protocols were: direct fine-tuning of OC20-pretrained checkpoints, cotuning those checkpoints with OC20 replay, and cotraining from scratch on mixed data from both AQCat25 and OC20.

It is important to note how baseline performance was assessed in this context: evaluations of pretrained models on AQCat25 used the provided structures directly, without re-optimizing lattice constants using OC20 DFT settings. These metrics represent the performance inherited for subsequent tuning rather than the inherent capability of the OC20 model on these specific materials had geometries been fully relaxed with consistent settings. Similarly, all evaluations performed on the OC20 validation subset utilized structures with OC20-optimized lattice constants.

\subsubsection*{Architecture}

\begin{wrapfigure}{r}{0.47\textwidth}
    \vspace{-25pt} 
    \centering
    \includegraphics[width=0.47\textwidth]{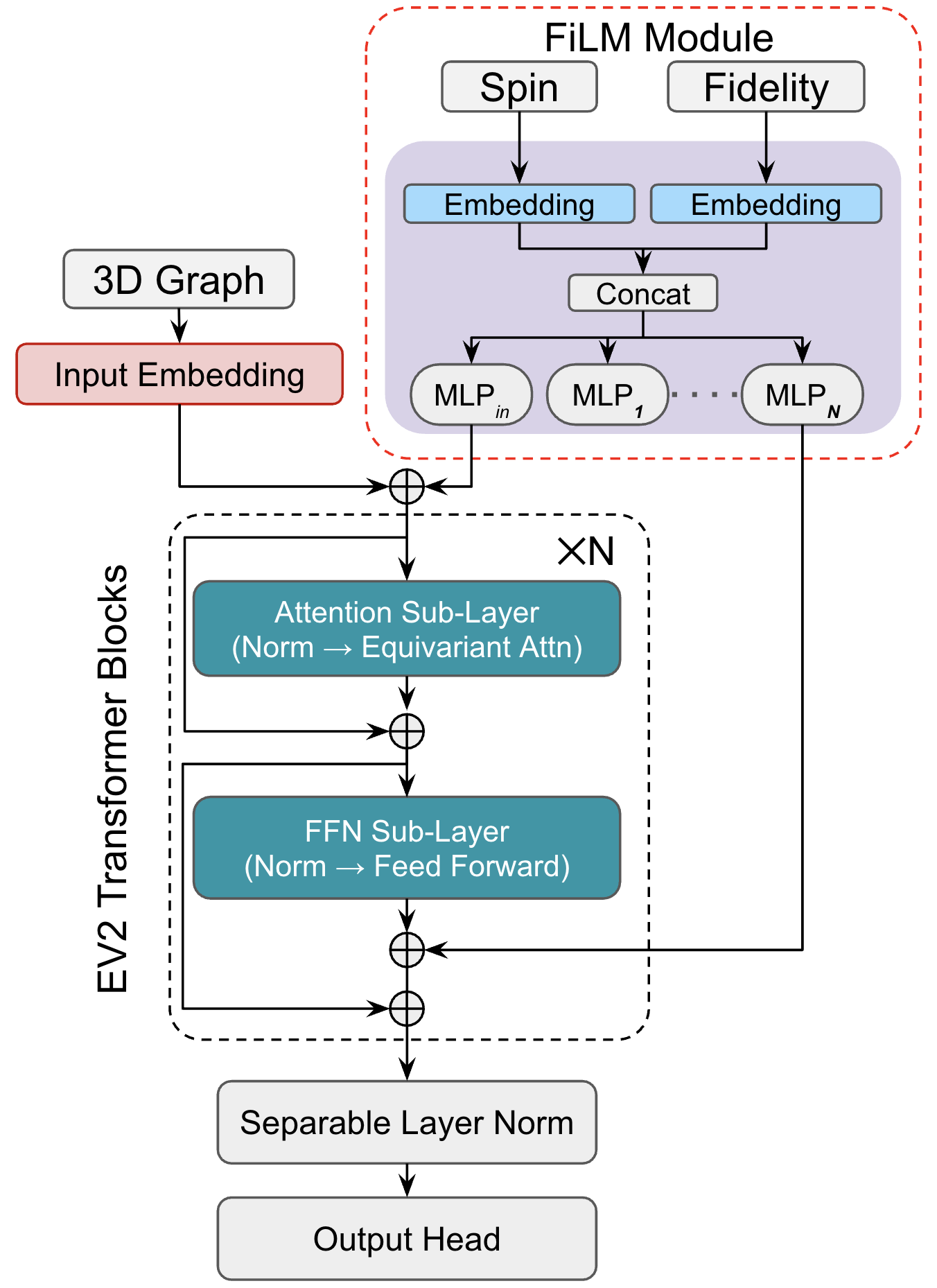}
    \captionof{figure}{FiLM module: binary context (spin, fidelity) $\rightarrow$ embeddings $\rightarrow$ MLP $\rightarrow \beta$; $\beta$ additively modulates scalar channels at the input (inFiLM) and optionally mid-block (in+midFiLM).}
    \label{fig:film-fig}
    \vspace{-10pt} 
\end{wrapfigure}

EV2 is an E(3)-equivariant transformer over atomic graphs: atoms form nodes, edges use pairwise distances and spherical harmonics, attention layers are equivariant, and feed-forward layers use $S^2$ activations \cite{cohen2018spherical, zitnick2022spherical, passaro2023reducing, liao2023equiformerv2}. Energies are predicted by a scalar head, and forces are predicted directly via a vector head. Architectural hyperparameters are listed in Table~\ref{tab:model_hyperparams}.

FiLM conditioning supplies the network with compact context about the \ac{DFT} settings of each structure. Two binary indicators encode spin treatment and fidelity (\texttt{spin\_on} and \texttt{low\_fi}). Each indicator is embedded; the embeddings are concatenated and passed through a small \ac{MLP} to produce a modulation vector $\beta$. We apply $\beta$ additively to the scalar channels broadcast across nodes (and per block for EV2-in+midFiLM). Preliminary tests with multiplicative factors $\gamma$ did not improve validation metrics measurably, so we retained the additive shift only for simplicity. Figure~\ref{fig:film-fig} diagrams the module and its insertion points.


\subsubsection*{Training and adaptation protocols}
\paragraph{Direct fine-tuning}

Starting from public EV2 OC20 All+MD checkpoints (31M and 153M parameters),  we fine-tuned on AQCat25 directly. These initial experiments tested the model's adaptation to the AQCat25 domain under distribution shift.

\paragraph{Cotuning with OC20 replay}
We fine-tuned the OC20 checkpoints on a composite stream consisting of AQCat25 high-fidelity and, when specified, a small AQCat25 low-fidelity spin-on stream, mixed with OC20 spin-off data at 0M/2M/20M scales.

\paragraph{Cotraining from scratch}
We trained EV2, EV2-inFiLM, and EV2-in+midFiLM from random initialization on the same composite streams used for cotuning.

\subsubsection*{Hyperparameters, controls, and compute}
Optimization and architectural settings are summarized in Tables~\ref{tab:model_hyperparams} and \ref{tab:training_hyperparams}. For each training run, 8$\times$H100 NVIDIA GPUs were used. Unless noted, the force term dominated the objective; the default loss ratio was $\lambda_E\!:\!\lambda_F=4\!:\!100$. To reduce the computational cost for the extensive model adaptation experiments, the AQCat25 dataset component was subsampled. Models were trained in single precision for a consistent comparison across the numerous experimental conditions and ablations. For tuning experiments, we tested stronger regularization by increasing weight decay and by lowering the learning rate; both choices produced early plateaus and higher validation errors on AQCat25 relative to fully thawed baselines. Except for models trained from scratch solely on AQCat25, energy and force targets were normalized using mean/standard deviation values from the OC20 distribution, as this yielded slightly improved performance in preliminary tests. We also tried incremental thawing schedules that kept the backbone frozen while adapting only input embeddings (to accommodate new elements), followed by gradual unfreezing. These schedules underperformed fully thawed tuning on AQCat25. 

Additionally, we explored alternative conditioning mechanisms and found that a simpler baseline involving direct concatenation of context embeddings performed competitively with FiLM when cotuning with limited (2M) OC20 data replay. We further experimented with more complex architectural modifications aimed at adapting the pretrained weights, including adding separate prediction heads routed by the conditioning flags and incorporating lightweight adapter modules within the transformer blocks. However, these approaches did not yield significant performance enhancements over the FiLM-based conditioning and fully thawed training strategies presented here. Finally, we do not claim hyperparameter or schedule optimality. Alternative warmup/decay, replay curricula, batch-composition policies, weight decay, EMA, or gradient clipping may yield further gains. Our goal here is a consistent and reproducible setup that enables clear comparison across regimes and architectures for the baseline models being presented.
\section*{Results and Discussion}
\subsection*{Dataset composition}
\begin{figure}[ht]
    \centering
    \includegraphics[width=\linewidth]{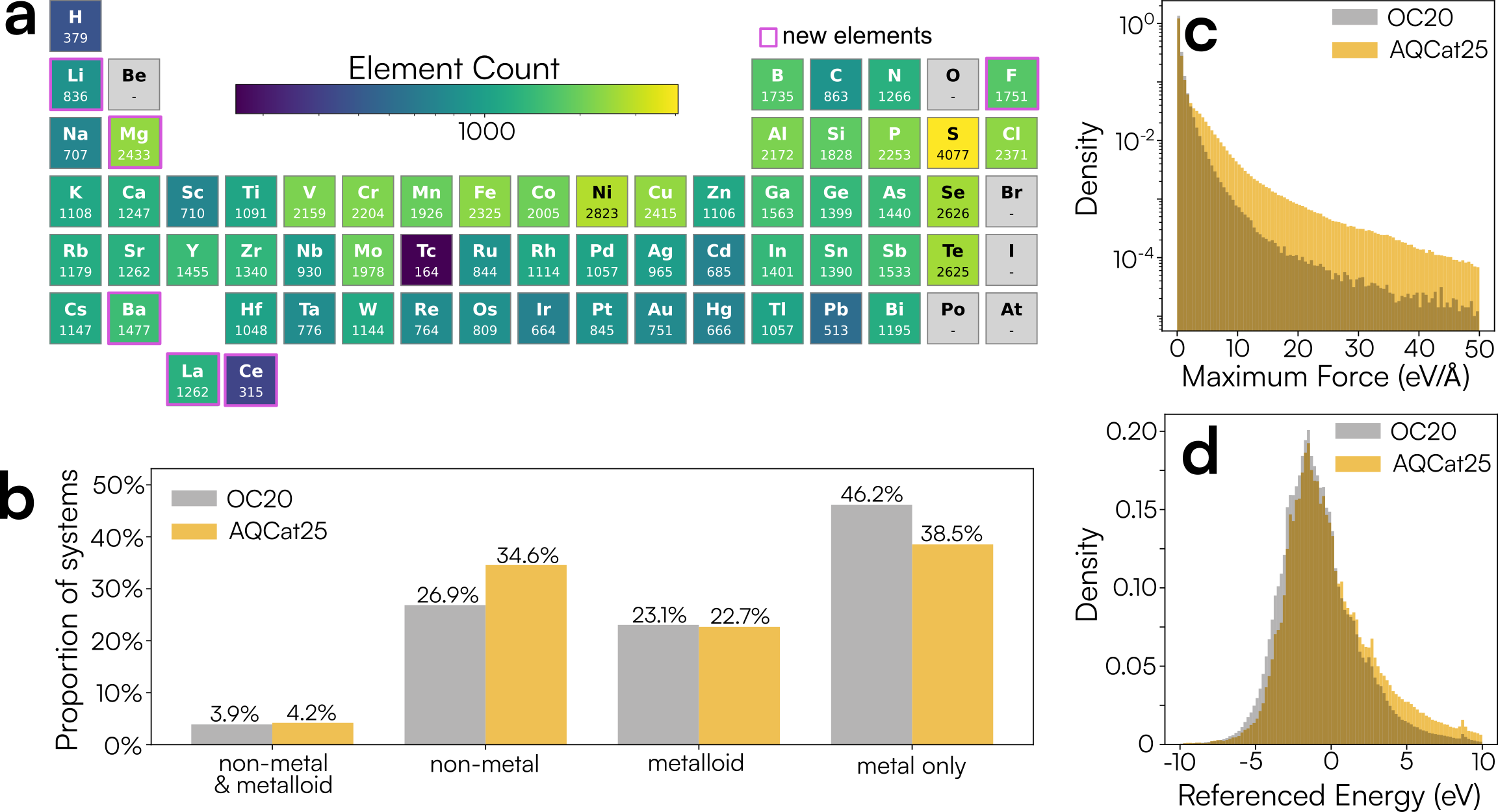}
    \caption{Summary statistics on the AQCat25 dataset and some comparisons to the \ac{OC20 dataset}\cite{chanussot2021open}. (a) Element counts showing the frequency with which each element appears in the dataset. (b) The proportion of systems that fit into four material-type categories for the entire training splits \ac{OC20 dataset} and AQCat25. (c \& d) The distribution of adsorption energies and maximum forces across the training split of AQCat25 and the 2M training subsplit of OC20.}
    \label{fig:dataset-figure}
\end{figure}

Some summary statistics about the AQCat25 dataset and how it compares to the \ac{OC20 dataset} are shown in Figure \ref{fig:dataset-figure}. As can be seen in Figure \ref{fig:dataset-figure}b, there is a significantly higher proportion of non-metal systems and lower proportion of metal systems in the AQCat25 dataset compared to the \ac{OC20 dataset}. The proportion of the other two categories, however, (non-metal \& metalloid and metalloid) are roughly equal. Because non-metal systems typically have poorer performance compared to intermetallics\cite{kolluru2022open}, we will look at key model performance metrics split over these material categories. The adsorption energy and maximum force distributions (Figure \ref{fig:dataset-figure}c-d) reveal that the AQCat25 dataset is biased towards higher force, higher energy systems when compared to OC20. This can be explained by the more aggressive approach taken when sampling high-force systems. Here, we used larger standard deviations to sample rattled configurations and also included the high energy transition state like systems.
\subsection*{Optimizing Data Generation Strategies for Fine-tuning}

\begin{figure}[ht]
    \centering
    \includegraphics[width=\linewidth]{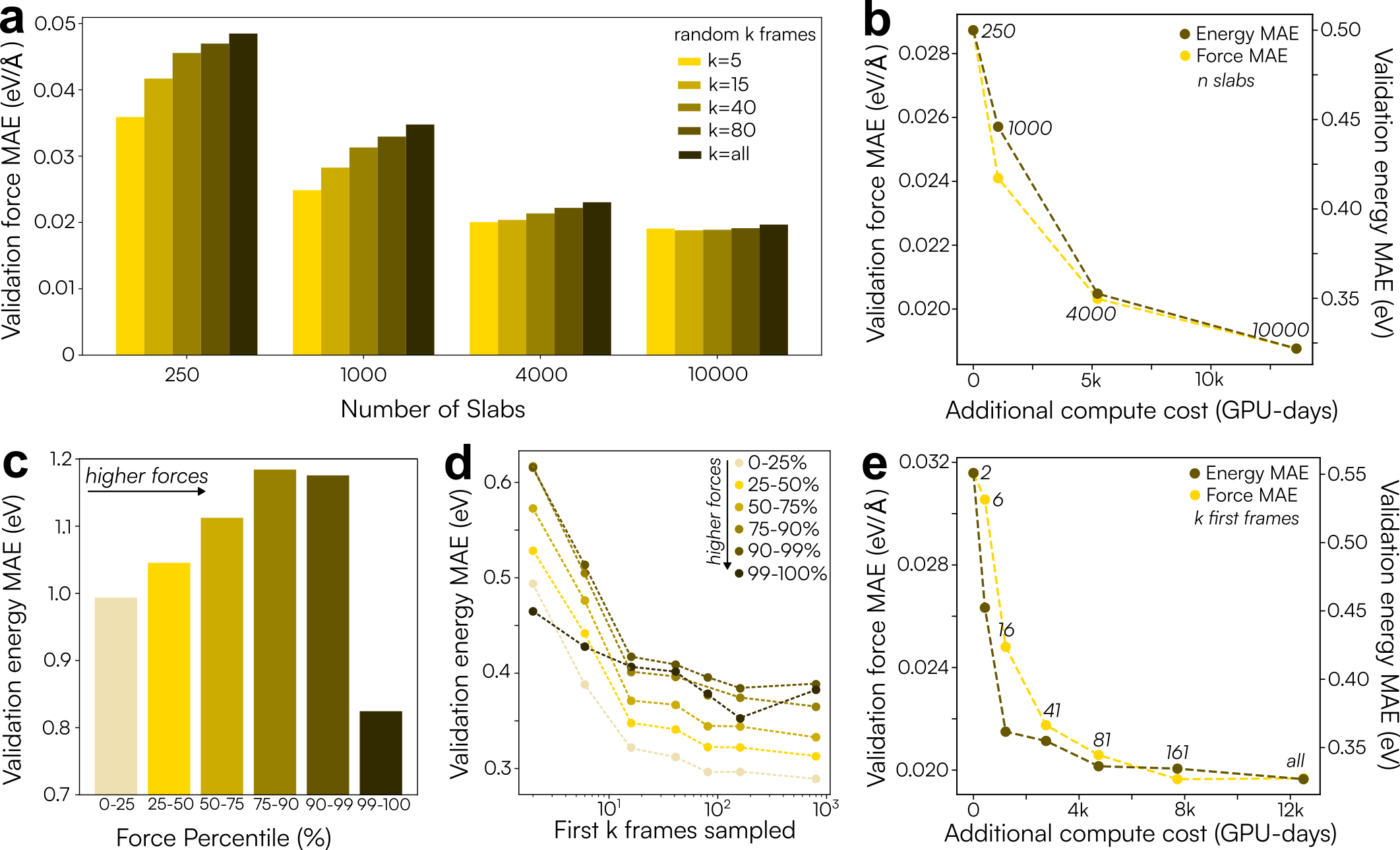}
    \caption{An exploration of opportunity for cost reduction in model training (a) and data generation (b-e) by directly fine-tuning 31M parameter Equiformer v2 models. (a) The validation force \ac{MAE} obtained when training models on \textit{random} subsamples of the data for four dataset sizes. (b) The incremental compute cost for increasing dataset size. (c) The pretrained 31M parameter Equiformer v2 model energy \ac{MAE} on force percentile segregated data from the AQCat25 validation dataset. (d) Evolution in force percentile segregated energy MAE for fine-tuned models trained on variable \textit{first k} frames from the dataset. (e) The trade-off between cost to generate training data and model performance when considering terminating relaxations after \textit{first k} steps.}
    \label{fig:subsampling}
\end{figure}

We wanted to explore opportunities to improve our data generation strategy to maximize model performance while minimizing the cost associated with dataset generation and model fine-tuning. To do this, we looked at the change in model performance as a function of two variables: number of \ac{DFT} single points seen per system and number of slabs. For heterogeneous catalyst systems, there are two types of diversity the models must generalize across: (1) the adsorbates and (2) the material surfaces the adsorbates are adsorbed to. The latter is much more complex. We initially adopted a scheme of performing four adsorbate-slab relaxations per slab to reduce the number of slab relaxations that needed to be performed as this data is not directly used in model training for referenced energy models. This turned out to be a suboptimal choice, however, because material diversity is important to tackle. For the future, we would choose to perform one adsorbate-slab calculation per slab. In Figure \ref{fig:subsampling}, we show the relationship between model force performance, number of slabs seen, and number of \ac{DFT} single points seen per system. Models were directly tuned starting from the publicly available 31M parameter EV2 model (All + MD) \cite{liao2023equiformerv2}.

\begin{figure}[ht]
    \centering
    \includegraphics[width=0.8\linewidth]{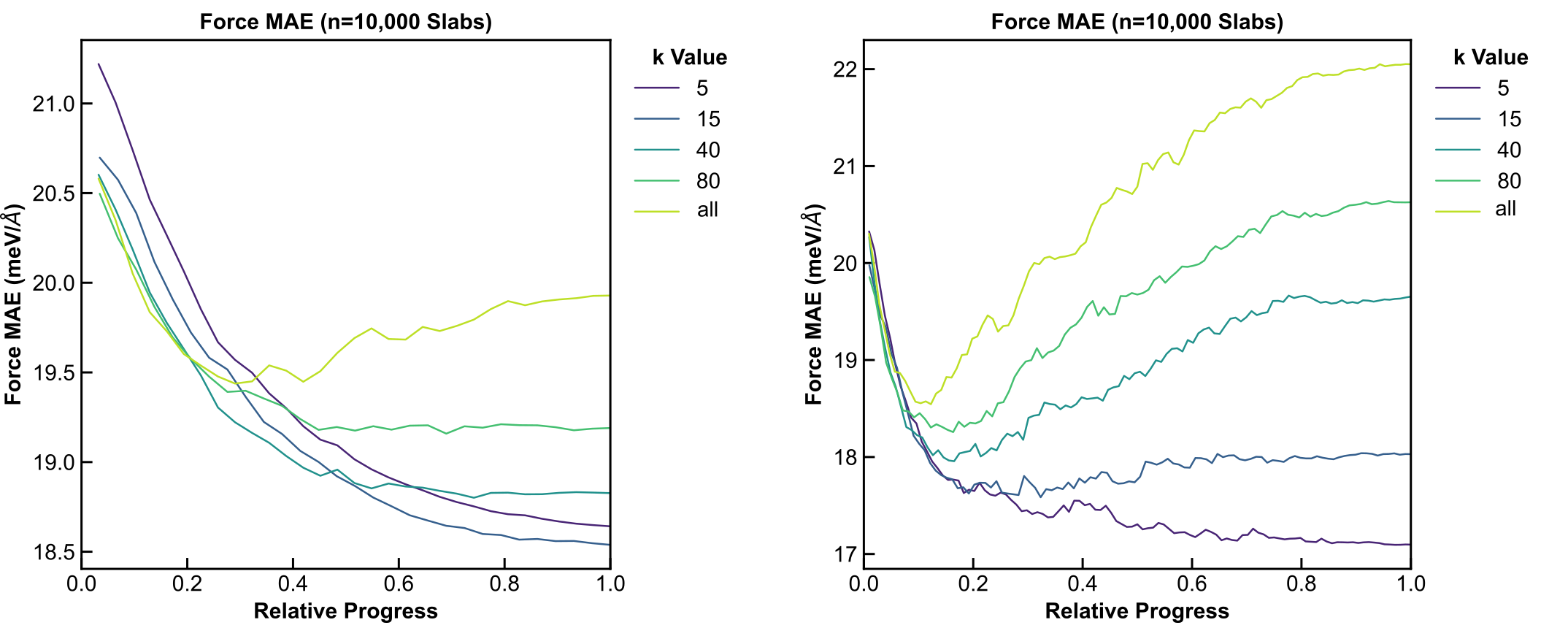}
    \caption{Model overfitting for direct tuning using the 31M parameter (left) and 153M parameter Equiformer v2 model. Sampling frames (as indicated by the k-values) reduces overfitting.}
    \label{fig:overfitting}
\end{figure}

The amount of data used to train the models directly impacts the cost to train and to iterate between architectures and ablations. Gasteiger and colleagues have shown the OC20-2M subset to be representative of the full OC20 dataset, primarily because it preserved the underlying chemical diversity\cite{gasteiger2022gemnetoc}. Other approaches use more complex stratified sampling (based on feature-space clustering) to ensure that diverse, high-energy, and uncommon configurations are explicitly captured to improve model robustness\cite{qi2023direct}. Given the precalculated training data, we explored the opportunity to reduce model training costs by sampling the frames along adsorbate-slab relaxation trajectories. Sampling was performed using force-stratified selections of the trajectory to obtain a representative distribution of systems. For consistency, models were trained for a nearly constant number of total gradient updates, approximately equivalent to the number of steps in one epoch of the largest split. Further, the data ablation for sampling frames from trajectories (Figure \ref{fig:subsampling}a) was designed to probe redundancy in highly autocorrelated data and thus only applied to the relaxations and \ac{MD} data; the rattled and transition state data were included entirely for each model. To enable a cleaner evaluation of the data cost-benefit trade-off for tuning, without confounding the results with the model's ability to learn new, unseen elements, we restricted the subsampling experiments to AQCat25 systems with elements already seen by the 31M pretrained model. Figure \ref{fig:subsampling}a reports the validation \ac{MAE} from the final training checkpoint, which highlights the risk of overfitting. Conversely, Figure \ref{fig:subsampling}b plots the MAE from the best-performing model during training for each slab count (averaged over all k values) against the data generation cost. As can be seen in Figure \ref{fig:subsampling}a, for small numbers of slabs, sampling a subset of frames rather than using the full trajectory actually leads to better force \ac{MAE} values. This is likely due to a high propensity for overfitting, which is shown in Figure \ref{fig:overfitting} and remedied by sampling. At larger slab numbers, the differences are minor but the cost to train will be lower if sampling is performed. For energy \ac{MAE} there is not a substantial trend with changing sample size (see Figure \ref{fig:subsampling-supplemental}). Therefore, sampling frames is a useful cost-saving strategy. We adopted a subsampling approach for the model adaptation experiments presented in subsequent sections. However, for those experiments, we employed random sampling per trajectory rather than the force-stratified approach. We found this yielded slightly improved performance, likely due to increasing the representation of low-force, near-equilibrium frames that are critical for downstream adsorption energy tasks.

Unsurprisingly, having more unique slabs in the dataset improves performance. However, there is a cost trade-off to be made, which is explored in Figure \ref{fig:subsampling}b. Comparing 250 to 1,000 and 1,000 to 4,000 there is a large improvement in the metrics. Going from 4,000 to 10,000 slabs, however, we are beginning to enter the domain of having diminishing returns on our computational investment, indicating that the number of slabs we calculated in this dataset was a reasonable choice. Nonetheless, this experiment primarily assessed convergence with respect to the number of unique slabs, not the total number of unique adsorbate-catalyst combinations, which warrants further investigation. Here, the additional compute cost is referenced to the 250-slab dataset. This value serves as a proxy for the total computational investment, which is expected to correlate with the true data generation cost and illustrates the trend of diminishing returns.

The apparent redundancy revealed by randomly sampling offers a potential opportunity: what if instead of optimizing full relaxation trajectories we instead only calculate the first k points? This could greatly reduce the compute cost to generate the data, but it introduces a potential new problem. It biases the relaxation data towards higher force states which could cause models trained on the data to have poor performance on low force systems. To investigate this, we divided the validation set into force percentiles and examined changes in performance on the different percentiles. As a baseline, we first assessed the pretrained model performance in Figure \ref{fig:subsampling}c. Performance decreases with increasing force percentile with the exception of the highest force percentile considered, which has better performance than even the lowest force percentile. This is likely a reflection of the underlying \ac{OC20 dataset} that contains, \ac{MD}, rattled systems, and relaxations. \ac{MD} and rattled data have high forces, while relaxations contain many frames in the low force regime. Figure \ref{fig:subsampling}d shows the evolution of model performance on the force percentile segmented validation split with an increasing number of frames sampled. Please note that here the data ablation also only applied to relaxations; all models were trained on the \ac{TS}-like and rattled data, but none included the \ac{MD} data. Performance overall improves with the number of frames sampled but it does not occur in a way that disproportionally affects specific force segments from 0-99\%. The one exception to this is the highest force percentile which modestly improves with increasing k. This is because all models used were trained using the very high force data (rattled and \ac{TS}). Performance in this percentile is most influenced by high force data. 

Using the first k relaxation frames presents an interesting trade-off between compute cost and model performance which is captured in Figure \ref{fig:subsampling}e. By only calculating between 40 and 80 frames instead of up to 800, we can achieve a Pareto optimum in model performance and compute cost for dataset generation. This would be our recommendation for future data generation campaigns. This exploration also revealed the advantage of training total energy models when designing a dataset to fine-tune models with cost in mind. If just 41 adsorbate-slab frames are computed, on average 75\% of the compute would be spent relaxing the slab completely. At 81 frames, this cost decreases to 63\%, but it is still substantial. For total energy models, this cost is not necessary because relaxed slab energies are not needed to train. We also recommend designing datasets to train total energy models for future campaigns.

\subsection*{Model Adaptation Strategies}\label{sec:cotuning_cotraining}
\begin{figure}[ht]
    \centering
    \includegraphics[width=\linewidth]{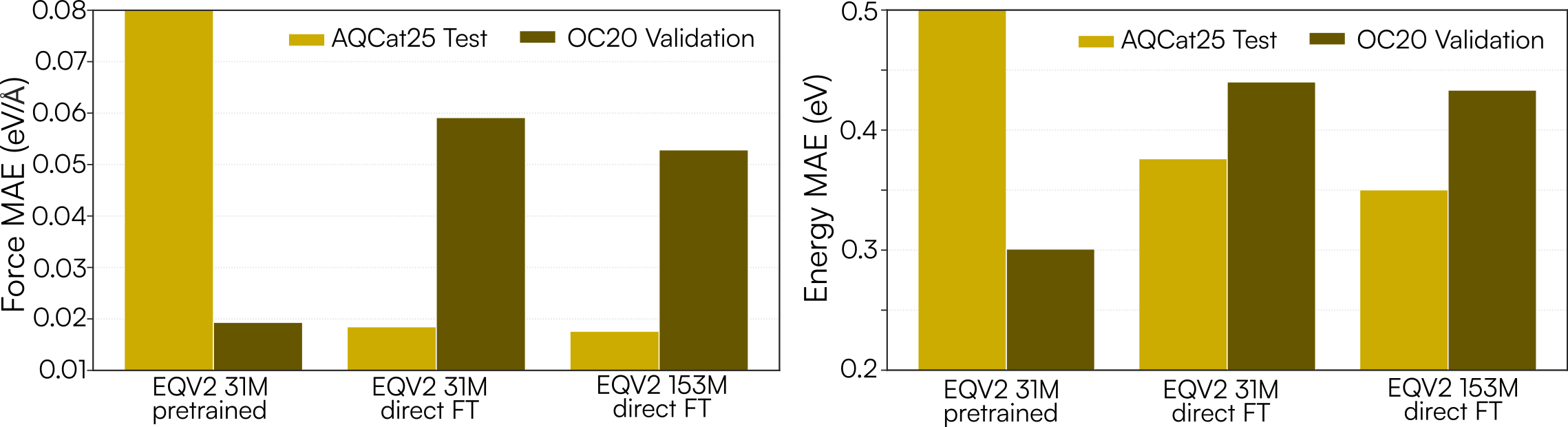}
    \caption{Test force and energy \ac{MAE} on AQCat25 test and a system-stratified subsample of the OC20 validation OOD both split for three different models: a pretrained (on OC20 only) 31M parameter EV2 model, a 31M parameter EV2 directly fine-tuned (FT) on AQCat25, and a 153M parameter EV2 directly fine-tuned on AQCat25}
    \label{fig:fine-tune-motivation}
\end{figure}

Although total energy models offer a clear path forward for more efficient data generation, this study focused on the adsorption energy target. Though arguably a more challenging learning task, as the model must implicitly account for the bare slab reference energy and any restructuring, adsorption energy may offer a significant convenience in established catalysis workflows. It also provides a well-defined target that isolates the adsorbate-surface interaction, which is ideal for developing a mixed fidelity/mixed physics adaptation protocol. Moreover, our overall objective is to create an MLIP that is broadly applicable, so we also wanted to understand model performance on the OC20 validation set. To assess the performance drift on the original OC20 task during these and subsequent experiments, we utilized a system-stratified subsample of the OC20 Val OOD Both split. This subset, which was sized to be comparable to individual AQCat25 validation splits ($\sim$300k frames), is a computationally efficient metric for relative comparisons between models. The resulting MAE values, however, may not reflect absolute performance on the full OC20 distributions. An evaluation of performance for a pretrained 31M parameter EV2 model and two directly fine-tuned (FT) EV2 models with 31M and 153M parameters on AQCat25 test and the subsampled OC20 validation split are shown in Figure \ref{fig:fine-tune-motivation}. We find that direct fine-tuning delivers reasonable AQCat25 errors, but deviation from the OC20 baseline on its validation split is significant. As anticipated, increasing model capacity from 31M to 153M parameters generally improves energy metrics on the AQCat25 test set. This is also true for increasing the energy loss weight ($\lambda_E$) (see Table~\ref{tab:model-performance2}). The 153M model with $\lambda_E=100$ yields the best energy \ac{MAE} metrics in these direct fine-tuning experiments (Table~\ref{tab:model-performance2}). However, the gains achieved by the larger 153M model may not justify its increased computational cost for practical applications, and this larger model still suffers from a significant performance drift for the original OC20 task.

\begin{figure}[ht]
    \centering
    \includegraphics[width=0.7\linewidth]{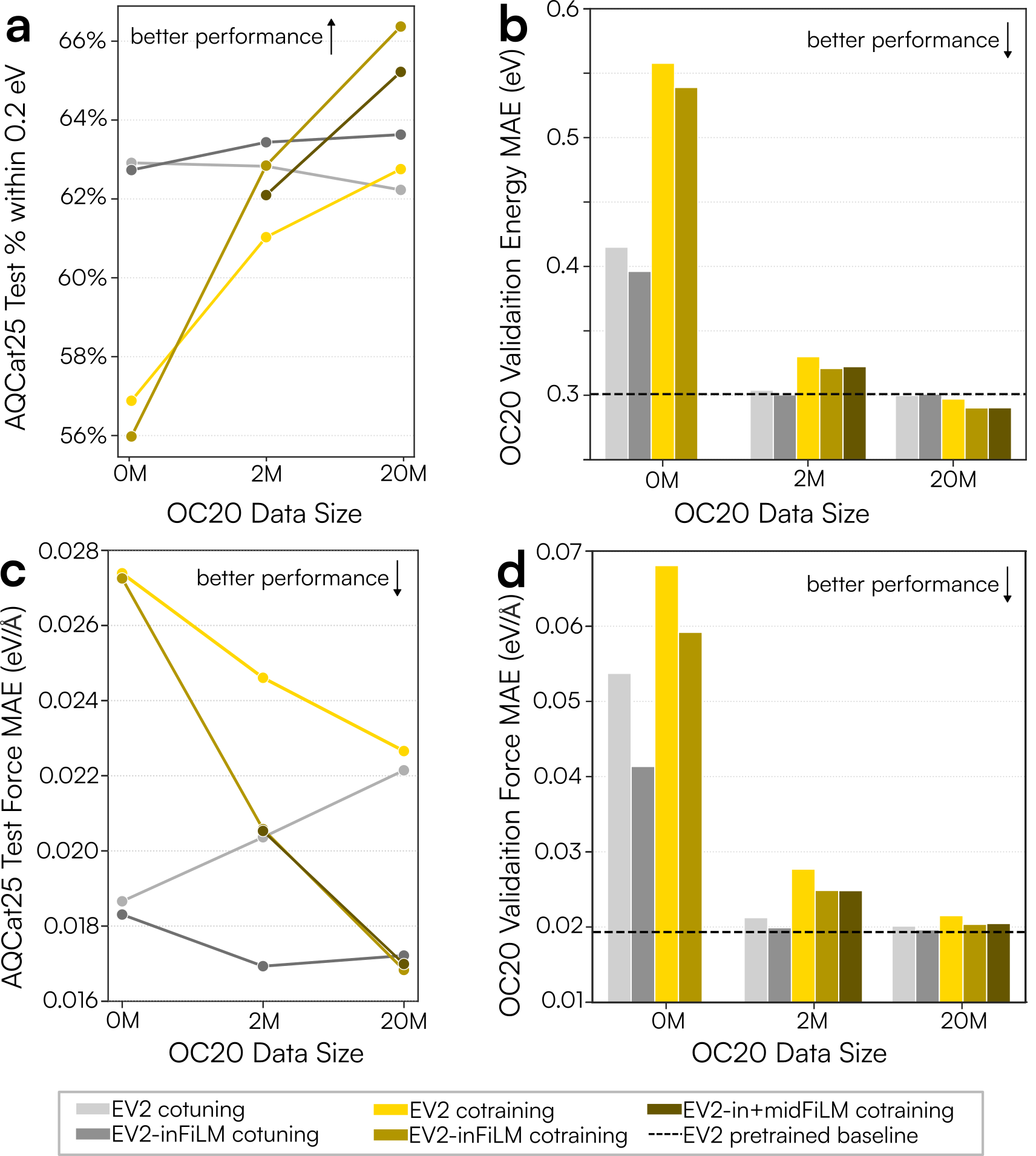}
    \caption{Model performance under cotuning and cotraining on AQCat25 and OC20. Part (a) shows the percentage of AQCat25 test set energies within 0.2 eV of the DFT value, while part (c) shows the force MAE on the AQCat25 test set, both as a function of the amount of OC20 data seen during training. Parts (b) and (d) explore the energy and force MAE trends on the OC20 validation set.}
    \label{fig:replay-figure}
\end{figure}

To mitigate this drift, we explored opportunities to cotune and cotrain models using both subsamples of the AQCat25 dataset and the \ac{OC20 dataset}. The models evaluated in this context, including those jointly trained with no additional OC20 data, incorporate the low-fidelity, spin-on data alongside the high-fidelity AQCat25 data. As seen in Figure~\ref{fig:replay-figure}b and d, we observe that 0M models (models trained without OC20 data) exhibit a substantial deviation from the baseline OC20 performance (dashed black line). Unsurprisingly, the exclusion of the spin-off OC20 dataset leads to poor performance on OC20 validation relative to baseline metrics, even with the inclusion of the small lowfi spin-on set. Therefore, to produce a model that performs well across all domains (high/low fidelity, spin on/off) within a practical model size, we investigate cotuning and cotraining (from scratch) strategies that incorporate two amounts of the original OC20 data. Figure~\ref{fig:replay-figure} summarizes the effect of including this OC20 data under two training regimes and three architecture variants.

Adding OC20 data consistently reduced deviation from the OC20 baseline on the examined validation split. For both cotuning and cotraining from scratch, the energy and force \ac{MAE} trend toward the baseline as the amount of OC20 data increases for both energies and forces (Fig. \ref{fig:replay-figure}b and d). We do not see this exact trend on the AQCat25 test split (Figure \ref{fig:replay-figure}a and c). In this case for energies we are showing the percent of frames that have an absolute energy error less than or equal to 0.2 eV. For this metric, a perfect model would have 100\%. This was done because we observed that the energy values had strong outliers. One group of systems contributing to this phenomenon are those where the slab is organic (entirely composed of non-metals). This approach as an alternative to \ac{MAE}, is an unbiased way to ensure strong outliers do not skew the results. We have included some additional Figures in the Supplementary Information to explore this metric further with different cutoffs (0.1, 0.3, 0.4, 0.6, and 0.8 eV instead of 0.2 eV) and using the \ac{MAE} for the energies instead on the test and validation splits. It seems as though the results are sensitive to this metric, so we will only make broad conclusions. The percentages of errors within the threshold for AQCat25 energies are largely unchanged when increasing data for cotuning, whereas with cotraining they increase (performance improves). For forces, there is a drastic increase in performance with more OC20 data for cotraining. For cotuning with FiLM there is a modest improvement in forces when including OC20 data, but a substantial degradation when cotuning without FiLM. An economic approach when considering cost to train and performance on both AQCat25 test and OC20 validation is achieved when cotuning with 2M OC20 examples. Cotraining from scratch with FiLM improves performance for systems that have higher errors though, so a tradeoff exists.

These patterns follow from standard behavior under distribution shift and multi-domain supervision. OC20 is broader and uses different \ac{DFT} settings than AQCat25. Fine-tuning only on AQCat25 moves the parameters toward that narrower distribution and forgets OC20-specific features. Adding replay during cotuning without FiLM counteracts forgetting but also pulls the solution toward OC20 conventions, which explains the performance degradation on AQCat25 forces in Figure \ref{fig:replay-figure}c. Introducing FiLM provides a framework to distinguish these distributions, which rectifies the decrease in the force metric. Starting from scratch changes the optimization path. The performance trends are strongly dependent on the OC20 data size used. Unsurprisingly, at 0M, jointly tuning clearly outperforms jointly training from scratch, but as OC20 data is added, their performance becomes sensitive to the metric. While tuning holds a slight advantage at a very strict 0.1 eV low-error threshold, the models cotrained from scratch show an advantage at higher cutoffs (e.g., 0.6-0.8 eV), indicating they are more effective at capturing outliers (see Figures \ref{fig:cotuning_metric_evo}a-f. FiLM makes the domain information explicit. Conditioning on spin and fidelity yields feature rescaling that reduces gradient interference between magnetic vs.\ non-magnetic and high- vs.\ low-fidelity cases.

\subsection*{Exploring robustness and generalization}
We also explored the robustness and generalizability of models to form a more complete assessment of model usability. To do this, we constructed an additional validation set aimed at assessing the ability of the models to identify the global minimum energy for a given adsorbate-slab combination in line with the approach presented by Lan et al\cite{lan2023adsorbml}. We further explored differences in model performance when segmenting the data by interesting splits, namely the material type (metal-only, non-metal, metalloid, and metalloid+non-metal), whether spin was on or off, and whether the elements in a material were all included in OC20 or not. 

\subsubsection*{Global minimum adsorption energy}
\begin{figure}[ht]
    \centering
    \includegraphics[width=\linewidth]{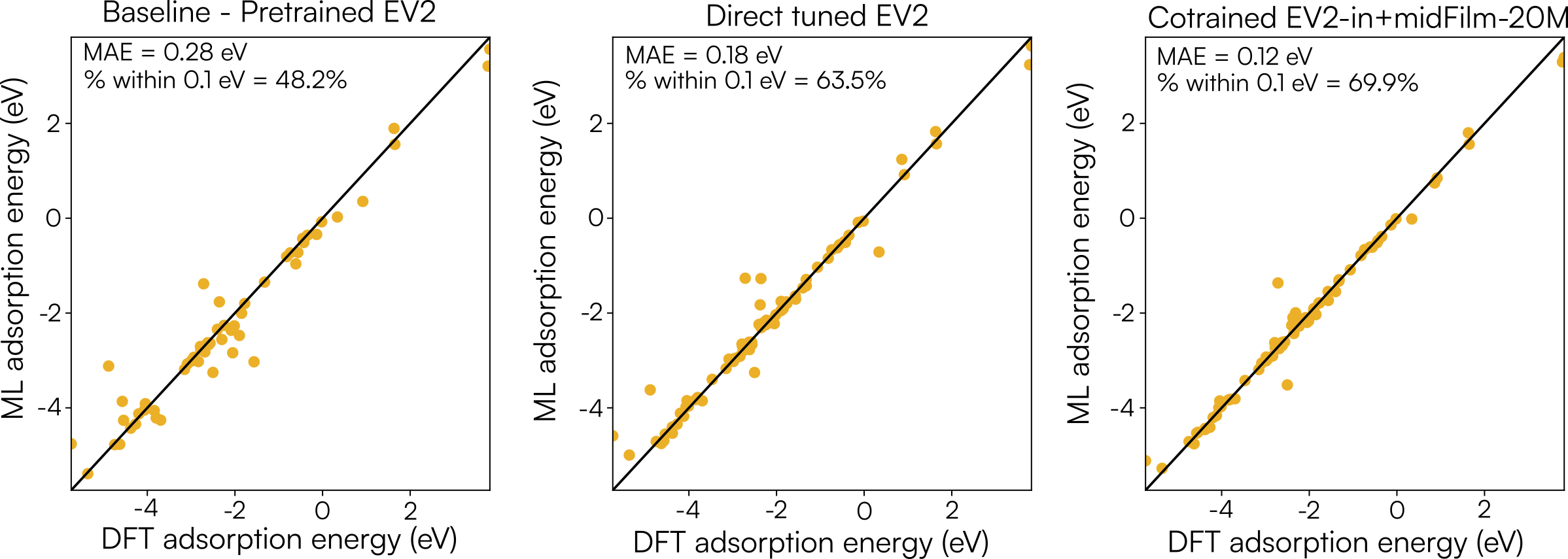}
    \caption{Parity plots between the DFT minimum adsorption energy and ML adsorption energy for a baseline pretrained 31M parameter EV2 model (left), the result of directly tuning that model on the AQCat25 dataset (center), and cotraining a 31M parameter EV2-in+midFilm model from scratch on both 20M examples from OC20 and the AQCat25 dataset (right).}
    \label{fig:adsorbml}
\end{figure}

The ultimate use of the \ac{MLIPs} trained here will be for practical catalyst discovery where an important figure of merit is the global minimum adsorption energy. We explored this using the dense \ac{DFT} validation set with 50 relaxations each for 109 adsorbate-slab combinations. We performed ML relaxation inference starting from the same initial configurations as \ac{DFT}. The relaxed states were filtered using the same algorithms presented by Lan et al.\cite{lan2023adsorbml} to find desorption, dissociation, intercalation, and significant surface change. Figure \ref{fig:adsorbml} compares the performance of three 31M parameter models on this task, using only the ML-predicted energies without DFT single-points on the ML-relaxed structures. On the left is the pretrained 31M EV2 model (trained on OC20 All+MD), taken from the publicly available fairchem checkpoint. For this model, inference on systems containing new elements were omitted since performance would be poor. In the center is a directly fine-tuned EV2 model. On the right is the EV2-in+midFilM model, which was cotrained using 20M OC20 examples and the AQCat25 dataset. As a point of comparison, when the OC20dense\cite{lan2023adsorbml} dataset was released, the EV2 model was not available. The best performing model was eSCN-MD-Large and on this task it had a 56.5\% success rate with an energy MAE of 0.17 eV\cite{lan2023adsorbml}. Here, success rate is defined as the percent of systems where the minimum adsorption energy found by ML is within 0.1 eV of the \ac{DFT} value. ML success metrics alone were included in a later release as 60.8\% for an EV2 model of unspecified size, 68.4\% for the UMA-S model, 71.1\% for the UMA-M model, and 74.4\% for the UMA-L\cite{wood2025family}. The success metric and MAEs have been annotated on the plots. We see the trend we would expect to see between models, with increasing performance from left to right. This further validates the usefulness of the models, and also supports the fact that the loss function and training metrics are well designed and correlate with our downstream use case.

\subsubsection*{Evaluating material and magnetic subsplits}

\begin{figure}[ht]
    \centering
    \includegraphics[width=\linewidth]{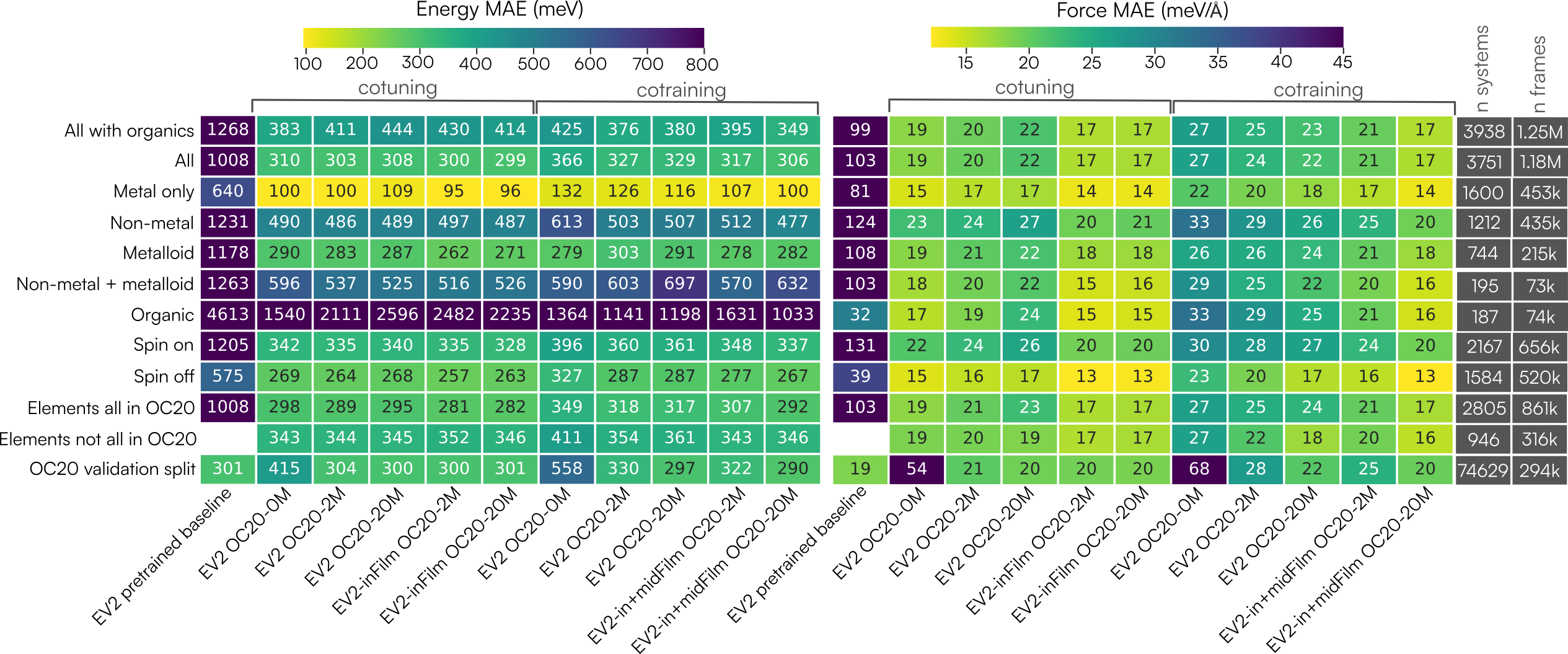}
    \caption{Comparison of Energy MAE (meV, left) and Force MAE (meV/Å, right) across different model training strategies and data subsets. Performance is evaluated for cotuning vs. cotraining, varying included OC20 data amounts (0M, 2M, 20M), and different architectures (vanilla EV2, EV2-inFiLM, EV2-in+midFiLM). Subsets include spin treatment, element novelty relative to OC20, and material type.}
    \label{fig:material_magnetism}
\end{figure}

To further probe the robustness of the models presented and identify potential systematic biases related to specific chemical or physical properties, we next evaluate performance across distinct subsets of the test data. Specifically, we analyze error trends based on the material type, whether the elements contained were all included in OC20, and whether spin polarization was treated during the calculation. The results of this are shown in Figure \ref{fig:material_magnetism} with model energy \ac{MAE} on the left and force \ac{MAE} on the right. The data presented here are evaluated on the AQCat25 test split for all rows except the OC20 validation split, which is the same subsample of the OC20 \ac{OOD} both split discussed above.

This analysis exposed a segment of the dataset that has very poor performance for energies: organic materials. Materials that only contain H, O, N, C, S, P, F, Cl, Br, I, and/or Se have very poor metrics as seen in the "Organics" row of Figure \ref{fig:material_magnetism}. This poor performance, however, does not extend to forces. This is likely because of the referencing scheme used. These materials are more able to restructure and it is therefore far more likely that the relaxed slab state is very different than the adsorbate-slab along the relaxation trajectory. These materials are not necessarily of catalytic interest, so it could be beneficial to be more selective when including them in the dataset. Certainly total energy models would be better suited to handle these systems because they remove dependence on the referenced state. Because of the significance of the errors on these systems, we removed them from all other splits presented in the figure. An alternative version of this figure has been included in the Supplementary Information where the organics are not segmented out. Without removing organics, we observe trends that are opposite to those expected and presented here.

 Here, we see that across the board model metrics for forces and energies are better for spin off than spin on. For most cases, when we compare the EV2 model to its corresponding EV2 + FiLM model, there is an improvement in spin on. Performance on systems where all elements appear in OC20 is better than systems that contain at least one new element. Aligned with existing precedent, the models more accurately predict energies and forces on metals when compared to other material types. Metalloids are slightly better treated than non-metals. Interestingly, energies for non-metals + metalloids are worse than non-metals, but the forces are not.

The models jointly trained from scratch outperform corresponding jointly tuned models by  $\sim$1 eV on the challenging organic material split, but achieve slightly worse performance on the other material splits excluding this category. The difference in optimization path and data exposure leads to these models marginally sacrificing performance on the broader set of materials to recover performance on this difficult class. Notably, as seen previously, the jointly trained from scratch model with 20M additional OC20 samples achieves the highest performance on the practical catalysis tasks described in Figures~\ref{fig:replay-figure}a and \ref{fig:adsorbml}.

A summary of model performance for the models discussed here and some others is shown in Table \ref{tab:model-performance}. Model names denote the architecture (EV2 31M default, inFiLM, or in+midFiLM) and training protocol, where 'ft' signifies fine-tuning an OC20-pretrained model and its absence means cotraining from scratch. Dataset identifiers specify OC20 data added (+OC20-2M/20M). Direct tuning experiments  excluded the low-fidelity spin-on subset.

 \begin{table}[ht]
    \centering
    \caption{Model Performance Metrics (Energy in meV, Forces in meV/\AA)}
    \label{tab:model-performance}
    \resizebox{\textwidth}{!}{%
    \begin{tabular}{l@{}lS[table-format=4.0]S[table-format=2.2]S[table-format=4.0]S[table-format=2.2]S[table-format=4.0]S[table-format=2.2]S[table-format=4.0]S[table-format=2.2]}
    \toprule
 Category & Model & \multicolumn{1}{c}{\makecell{All \\ E-MAE}} 
   & \multicolumn{1}{c}{\makecell{All \\ F-MAE}} 
   & \multicolumn{1}{c}{\makecell{Spin On \\ E-MAE}} 
   & \multicolumn{1}{c}{\makecell{Spin On \\ F-MAE}} 
   & \multicolumn{1}{c}{\makecell{Spin Off \\ E-MAE}} 
   & \multicolumn{1}{c}{\makecell{Spin Off \\ F-MAE}} 
   & \multicolumn{1}{c}{\makecell{OC20 Val \\ E-MAE}} 
   & \multicolumn{1}{c}{\makecell{OC20 Val \\ F-MAE}} \\
    \midrule
     Pretrained &  EV2-OC20 & 1268 & 98.63 & 1205 & 131.40 & 1378 & 37.06 & 301 & 19.32 \\
    \midrule
    \multirow[t]{2}{*}{Direct Tuning \hspace{10pt}} & EV2-OC20-ft-AQCat25-highfi only & 376 & 18.46 & 337 & 21.63 & 419 & 14.80 & 440 & 59.13 \\
     & EV2-OC20-ft-AQCat25-highfi only (153M) & 350 & 17.59 & 339 & 20.41 & 362 & 14.34 & 433 & 52.84 \\
    \cline{1-10}
    \multirow[t]{7}{*}{Cotuning} & EV2-OC20-ft-AQCat25 & 383 & 18.65 & 342 & 21.81 & 428 & 15.00 & 415 & 53.73 \\
     & EV2-inFiLM-OC20-ft-AQCat25 & 379 & 18.30 & 346 & 21.23 & 415 & 14.91 & 396 & 41.34 \\
     & EV2-OC20-ft-AQCat25+OC20-2M & 411 & 20.36 & 335 & 23.71 & 495 & 16.48 & 304 & 21.23 \\
     & EV2-inFiLM-OC20-ft-AQCat25+OC20-2M & 430 & 16.93 & 335 & 19.90 & 536 & 13.49 & 300 & 19.90 \\
     & EV2-OC20-ft-AQCat25+OC20-20M & 444 & 22.14 & 340 & 26.30 & 559 & 17.34 & 300 & 20.12 \\
     & EV2-inFiLM-OC20-ft-AQCat25+OC20-20M & 414 & 17.21 & 328 & 20.28 & 510 & 13.66 & 301 & 19.62 \\
     & EV2-inFiLM-OC20-ft-AQCat25+OC20-20M ($\lambda_E=100$) & 412 & 21.03 & 325 & 24.32 & 508 & 17.22 & 289 & 21.79 \\
    \cline{1-10}
    \multirow[t]{7}{*}{Cotraining} & EV2-AQCat25 & 425 & 27.38 & 396 & 29.86 & 457 & 24.53 & 558 & 68.06 \\
     & EV2-AQCat25+OC20-2M & 376 & 24.60 & 360 & 28.01 & 394 & 20.68 & 330 & 27.69 \\
     & EV2-inFiLM-AQCat25+OC20-2M & 392 & 20.57 & 345 & 23.79 & 442 & 16.86 & 321 & 24.86 \\
     & EV2-in+midFiLM-AQCat25+OC20-2M & 395 & 20.53 & 348 & 23.75 & 447 & 16.80 & 322 & 24.83 \\
     & EV2-AQCat25+OC20-20M & 380 & 22.65 & 361 & 26.85 & 402 & 17.81 & 297 & 21.51 \\
     & EV2-inFiLM-AQCat25+OC20-20M & 367 & \bfseries 16.83 & 334 & 19.90 & 403 & 13.28 & 290 & 20.35 \\
     & EV2-in+midFiLM-AQCat25+OC20-20M & \bfseries 349 & 16.98 & 337 & 20.05 & 363 & 13.44 & 290 & 20.46 \\
    \bottomrule
    \end{tabular}%
    }
\end{table}


\section*{Conclusion}

This work tackled a significant gap limiting the application of large-scale \ac{MLIPs} in heterogeneous catalysis: the proper treatment of magnetism and enhanced electronic fidelity to accurately model and discover novel catalysts containing earth-abundant, spin-polarized elements such as Fe, Co, and Ni. We demonstrated that while direct fine-tuning of a pretrained OC20 model on AQCat25 provides performance on the new data it leads to a significant degradation of performance on the original OC20 domain. We found that by combining the targeted high-fidelity physics captured in AQCat25 with the extensive chemical and structural diversity present in a large portion of the OC20 data, jointly training successfully enhances accuracy on the AQCat25 test set while mitigating degradation on the evaluated OC20 validation metrics. We further confirmed the applicability of our models for the practical catalysis task of identifying the global minimum adsorption energy on a diverse set of surfaces. This training methodology, utilizing multi-fidelity data and explicit conditioning, offers a promising path toward practical and broadly applicable \ac{MLIPs} for heterogeneous catalysis.
\section*{Contributions}
O.A.: Model implementation, experimental design, training, ablations, idea conceptualization, data processing, data analysis, data visualization, writing, editing. 
B.W.: Dataset generation, idea conceptualization, data analysis, data visualization, writing, editing. 
A.R.S.: Project leadership, idea conceptualization, writing, editing.
S.K., R.P., P.A., A.N.: Computing infrastructure.
T.S., J.C., T.L., R.W., S.A., A.R., A.F., C.W., A.W., T.M., K.R.: Code validation, manuscript review.

\section*{Acknowledgements}
The authors acknowledge SandboxAQ leadership, especially Ang Xiao, Adam Lewis, Arman Zaribafiyan, Jeff Graf, Takeshi Yamazaki, Nadia Harhen, Andrew McLaughlin, and Jack Hidary, for their support of this research. The authors thank Joseph Gauthier and Jens N{\o}rskov for valuable discussions and feedback. The authors recognize computational and engineering support from the Nvidia DGX team.

\bibliographystyle{unsrtnat} 
\bibliography{bib}

\clearpage 
\section*{Supplementary Information}

\subsection*{VASP parameters}
The \ac{VASP} parameters are summarized in Tables \ref{tab:vasp-pars} and \ref{tab:vasp-pars-md}. The Bloch vectors (kpoints) were set using the lattice vectors using the same technique implemented in the fairchem repository for slabs and adsorbate-slab systems. The z-direction is set to 1, while x and y are set using Equation \ref{eqn:kpoint}. For bulks, this calculation was also applied to the z-direction. For systems containing Ce, Co, Cr, Cu, Fe, Mn, Mo, Ni, Os, Ru, V, or W, spin polarization was enabled to account for magnetic effects.

\begin{equation}\label{eqn:kpoint}
    k = max\left[ \nint{\frac{40}{c}}, 1\right]
\end{equation}

\begin{table}[H]
    \centering
    \begin{tabular}{c|c|c}
       Variable & Setting Slabs Systems & Setting Bulks \\ \hline
        IBRION   & 2 & 1 \\
        NSW   & 800 & 250   \\
        ISIF   & 0 & 7 \\
        ISPIN   & 1 or 2 & 1 or 2 \\
        ISYM & 0 & 0 \\
        ALGO & Normal & Normal \\
        ISMEAR & 0 & 0 \\
        SIGMA & 0.1 & 0.1 \\
        EDIFFG & -0.03 & 1E-5 \\
        ENCUT & 500 & 500 \\
        PREC & Accurate & Accurate \\
        POTIM & 0.5 & 0.5 \\
        NELM & 250 & 250 \\
        EDIFF & 1E-4 & 1E-4 \\
        SYMPREC & 1E-10 & 1E-5 \\
        LREAL & Auto & False
        
    \end{tabular}
    \caption{\ac{VASP} parameters.}
    \label{tab:vasp-pars}
\end{table}

\begin{table}[H]
    \centering
    \begin{tabular}{c|c}
       Variable & Setting \\ \hline
        TEBEG   & 900  \\
        TEEND   & 900    \\
        MDALGO   & 1  \\
        ANDERSEN PROB   & 0.0  \\
        NSW & 80  \\
        POTIM & 2  \\
        IBRION & 0 \\
        NELMIN & 4 \\

    \end{tabular}
    \caption{\ac{MD} specific \ac{VASP} parameters.}
    \label{tab:vasp-pars-md}
\end{table}

\clearpage

\subsection*{Adsorbate referencing}

The adsorbate gas phase references were constructed using the energies of CO, H$_2$, H$_2$O, and N$_2$. Calculations were performed with the molecules separately in vacuum cubes of 10, 20, and 30 \AA. There was not a significant energy difference between 20 and 30\AA, so 30\AA~ was taken to be converged. The resulting per atom/element energies are summarized in Table \ref{tab:adsorbate-corrections}. 

\begin{table}[H]
    \centering
    \begin{tabular}{c|c}
       Atom & Energy [eV] \\ \hline
        H   & -3.4944 \\
        O   & -7.1590 \\
        C   & -7.2654 \\
        N   & -8.1351 
    \end{tabular}
    \caption{Adsorbate per atom energy corrections.}
    \label{tab:adsorbate-corrections}
\end{table}

\clearpage

\subsection*{Element counts OC20 versus AQCat25}
\begin{figure}[ht]
    \centering
    \includegraphics[width=0.8\linewidth]{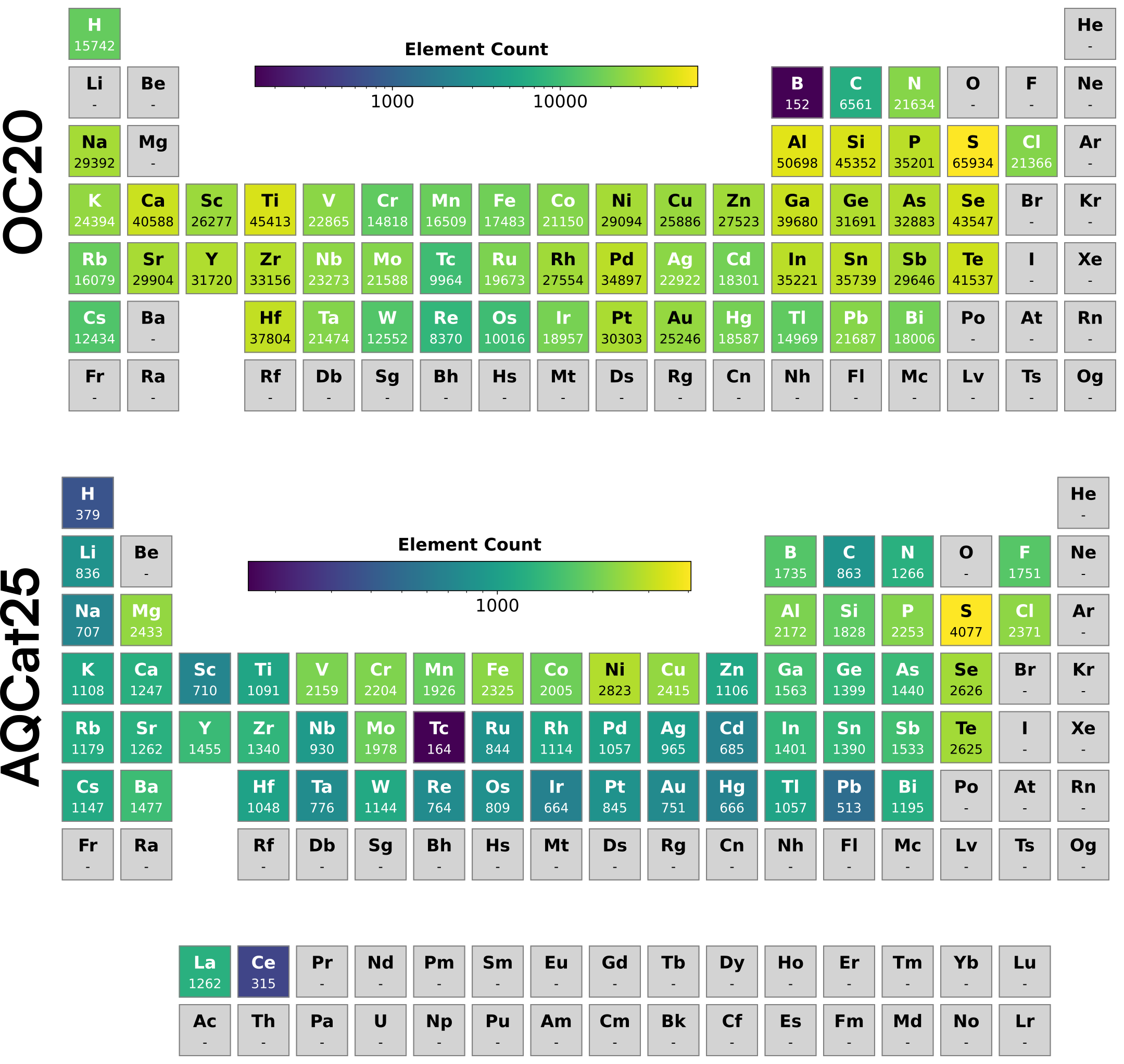}
    \caption{Element counts for all of the train splits of OC20 and AQCat25.}
    \label{fig:element-counts}
\end{figure}

Figure \ref{fig:element-counts} shows a comparison between the frequency with which elements occur in the AQCat25 and OC20 datasets. There are some notable differences like the presence of the six additional elements in AQCat25, the higher relative presence of boron in AQCat25, and the lower presence of Tc in AQCat25.

\clearpage

\subsection*{Sampling}

\begin{figure}[ht]
    \centering
    \includegraphics[width=0.7\linewidth]{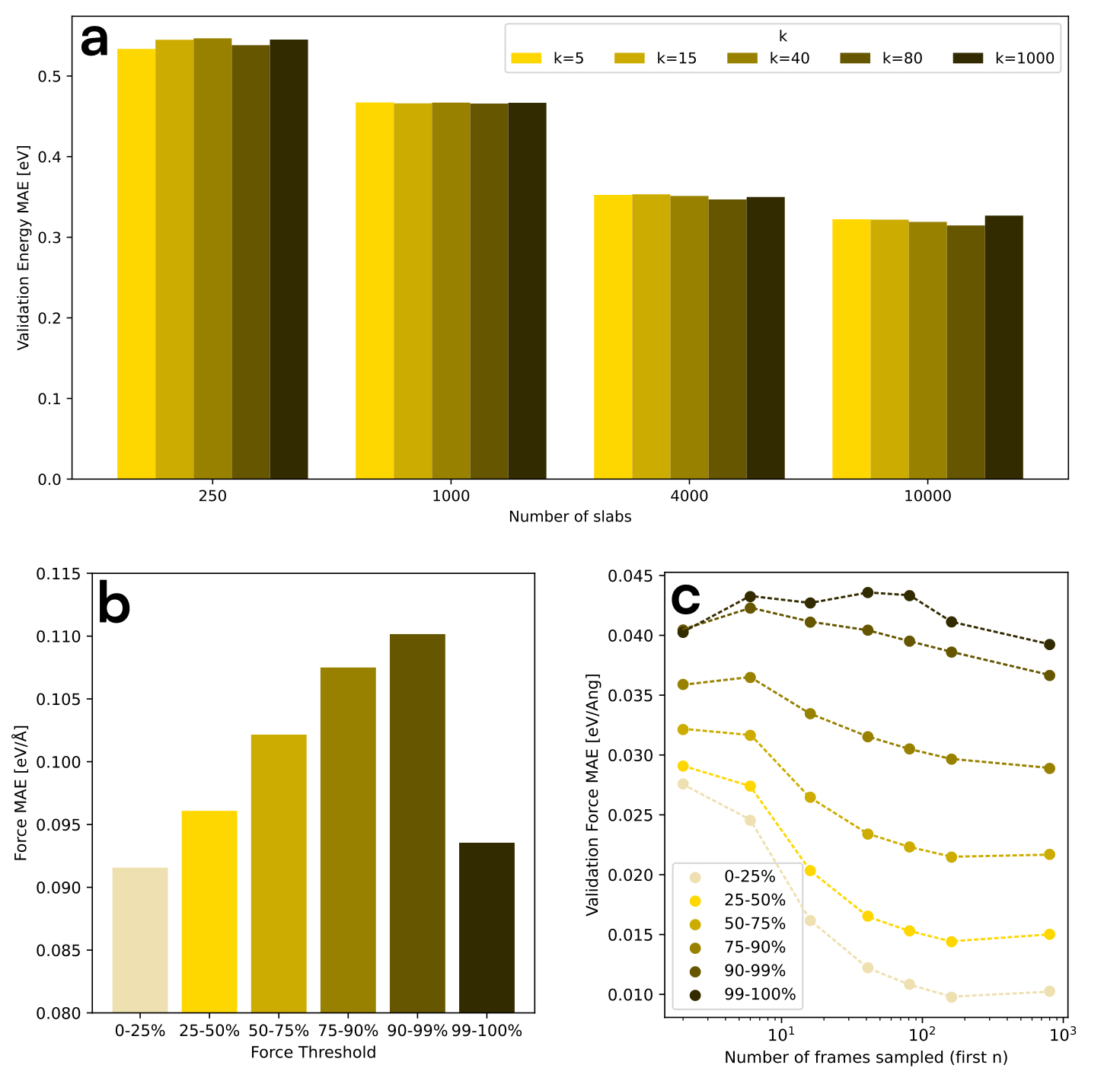}
    \caption{Complementary parts to the figure describing subsampling in the main text.} 
    \label{fig:subsampling-supplemental}
\end{figure}

Figure \ref{fig:subsampling-supplemental} shows: (a) Energy \ac{MAE} which does not show a substantial trend with changing k for random subsampling, but is improved by increasing the number of slabs. (b) The AQCat25 validation force \ac{MAE} for the pretrained 31M parameter Equiformer v2 model across stratified force bins, which shows roughly the same trend as energy: performance decreases on higher force systems with the exception of very high force frames which have better performance. (c) The AQCat25 validation force \ac{MAE} for naively finetuned models using different values of first k samples of the AQCat25 dataset to fine-tune.

\clearpage
\subsection*{Additional direct tuning metrics}
\begin{table}[H]
    \centering
    \caption{Model performance metrics for direct tuning (energy in meV, forces in meV/\AA)}
    \label{tab:model-performance2}
    \resizebox{\textwidth}{!}{%
    \begin{tabular}{l@{}lS[table-format=4.0]S[table-format=2.2]S[table-format=4.0]S[table-format=2.2]S[table-format=4.0]S[table-format=2.2]S[table-format=4.0]S[table-format=2.2]}
    \toprule
 & & \multicolumn{1}{c}{All E-MAE} & \multicolumn{1}{c}{All F-MAE} & \multicolumn{1}{c}{Spin On E-MAE} & \multicolumn{1}{c}{Spin On F-MAE} & \multicolumn{1}{c}{Spin Off E-MAE} & \multicolumn{1}{c}{Spin Off F-MAE} & \multicolumn{1}{c}{OC20 Val E-MAE} & \multicolumn{1}{c}{OC20 Val F-MAE} \\
Category & Model &  &  &  &  &  &  &  &  \\
    \midrule
    \multirow[t]{3}{*}{31M} & $\lambda_E=4$ & 376 & 18.46 & 337 & 21.63 & 419 & 14.80 & 440 & 59.13 \\
     & $\lambda_E=100$ & 372 & 20.48 & 349 & 23.90 & 398 & 16.52 & 458 & 57.67 \\
     & $\lambda_E=100$, with lowfi spin-on & 383 & 18.65 & 342 & 21.81 & 428 & 15.00 & 415 & 53.73 \\
    \cline{1-10}
    \multirow[t]{2}{*}{153M} & $\lambda_E=4$ & 350 & 17.59 & 339 & 20.41 & 362 & 14.34 & 433 & 52.84 \\
     & $\lambda_E=100$ & 343 & 19.71 & 335 & 22.77 & 352 & 16.16 & 420 & 47.07 \\
    \bottomrule
    \end{tabular}%
    }
\end{table}

\clearpage
\subsection*{Effect of toggling fidelity and spin flags}
\begin{figure}[ht]
    \centering
    \includegraphics[width=0.7\linewidth]
    {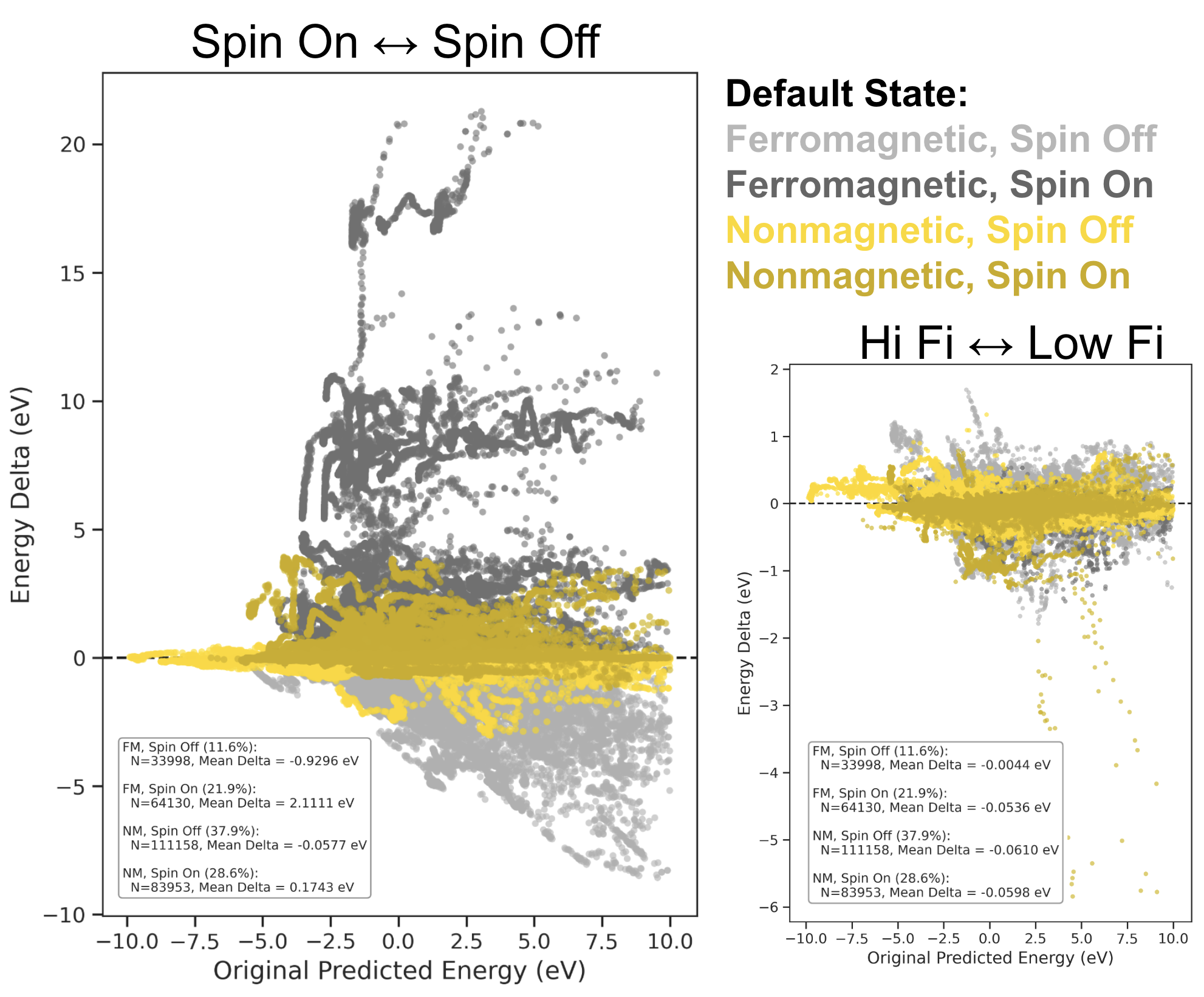} 
    \caption{Effect of toggling conditioning flags during inference on predicted energy delta ($\Delta E$). The left panel shows the impact of switching between \texttt{spin\_on} and \texttt{spin\_off} flags, while the right panel shows switching between high and low fidelity flags. Colors distinguish between ferromagnetic (FM, grey) and nonmagnetic (NM, yellow) systems based on their ground truth magnetic state (from the MP) and spin treatment in the dataset. Toggling the spin flag has a much larger effect on FM systems, including FM systems labeled as spin-off in the training data.}
    \label{fig:flag-toggling-ablation}
\end{figure}

 We also wanted to ablate the impact of the spin and fidelity flags on the resultant energies. Toggling the spin flag induces large energy shifts in opposite directions for systems categorized as ferromagnetic by the MP, depending on their original spin treatment: destabilizing correctly labeled spin-on systems (energy increases) and stabilizing incorrectly labeled spin-off systems (energy decreases). In contrast, NM systems show minimal energy changes when the spin flag is toggled, indicating the model correctly associates strong spin effects primarily with the FM materials (even those that excluded the elements that we categorized as necessitating spin treatment). Further analysis is needed to fully validate that the model has learned the correct underlying physics across domains. For instance, the observed asymmetry could simply reflect that evaluating FM systems with spin turned off represents a significant deviation from the training data distribution. The model may underperform in this regime because it has primarily learned patterns associated with spin-polarized FM states and lacks sufficient training examples or capacity to accurately model the less common or physically distinct spin-unpolarized state for these materials. 
 
\clearpage

\subsection*{Model Training Parameters}
\begin{table}[H]
\centering
\caption{Architectural hyperparameters for the EquiformerV2 models, including FiLM specifics. We refer to the EquiformerV2 paper \cite{liao2023equiformerv2} for a complete description of all architectural components, including normalization and activation functions.}
\label{tab:model_hyperparams}
\begin{tabular}{@{}ll@{}}
\toprule
\textbf{Hyperparameter} & \textbf{Value} \\
\midrule
\multicolumn{2}{l}{\textbf{Core EquiformerV2 Architecture}} \\
Number of Transformer blocks & 8 (31M), 20 (153M) \\
Embedding dimension $d_{embed}$ & 128 \\
$f_{ij}^{(L)}$ dimension $d_{attn\_hidden}$ & 64 \\
Hidden dimension in feed forward networks $d_{ffn}$ & 128 \\
Number of attention heads & 8 \\
Maximum spherical harmonic degree ($L_{max}$) & 4 (31M), 6 (153M) \\
Maximum spherical harmonic order ($M_{max}$) & 2 (31M), 3 (153M) \\
Dropout rate & 0.1 \\
Stochastic depth & 0.1 \\
Cutoff radius ($ \text{\AA} $) & 12.0 \\
Maximum number of neighbors & 20 \\
\midrule
\multicolumn{2}{l}{\textbf{FiLM Architecture Addendum (EV2-FiLM)}} \\
Auxiliary feature embedding dimension & 16 \\
MLP hidden dimension for modulation & 128 \\
MLP dropout & 0.1 \\
FiLM modulation strategy & \begin{tabular}[t]{@{}l@{}}
                             \textbf{Cotuning}: Input layer only \\
                             \textbf{Training from Scratch}: \\
                             \quad - Input layer only \\
                             \quad - Input layer \& all Transformer blocks
                             \end{tabular} \\
\bottomrule
\end{tabular}
\end{table}
\begin{table}[H]
\centering
\caption{Training and optimization hyperparameters for each experimental strategy.}
\label{tab:training_hyperparams}
\begin{tabular}{@{}llll@{}}
\toprule
\textbf{Parameter} & \textbf{Direct Finetuning} & \textbf{Cotuning} & \textbf{Training from Scratch} \\
\midrule
Pre-trained Checkpoint & OC20 All+MD & OC20 All+MD & None \\
\midrule
\multicolumn{4}{l}{\textbf{Optimizer}} \\
Optimizer & \texttt{AdamW} & \texttt{AdamW} & \texttt{AdamW} \\
Weight decay & $1 \times 10^{-3}$ & $1 \times 10^{-3}$ & $1 \times 10^{-3}$ \\
Learning rate (31M) & $7 \times 10^{-5}$ & $7 \times 10^{-5}$ & $4 \times 10^{-4}$ \\
Learning rate (153M) & $8 \times 10^{-5}$ & $8 \times 10^{-5}$ & $4 \times 10^{-4}$ \\
LR scheduling & \multicolumn{3}{c}{Cosine annealing with linear warmup} \\
Warmup epochs & 0.01 & 0.01 & 0.1 \\
Model EMA decay & 0.999 & 0.999 & 0.999 \\
\midrule
\multicolumn{4}{l}{\textbf{Batch Size \& Epochs}} \\
Batch size per GPU (31M) & 20 & 20 & 20 \\
Batch size per GPU (153M) & 6 & 6 & 6 \\
Gradient accumulation (153M) & 3 steps & 3 steps & 3 steps \\
Effective batch size (31M) & 160 & 160 & 160 \\
Effective batch size (153M) & 144 & 144 & 144 \\
Max epochs & 30 & 30 & 30 \\
\midrule
\multicolumn{4}{l}{\textbf{Loss \& Regularization}} \\
Energy coefficient ($\lambda_E$) & 4, 100 & 4 & 4 \\
Force coefficient ($\lambda_F$) & 100 & 100 & 100 \\
Gradient clipping norm threshold & 5 & 5 & 100 \\
\bottomrule
\end{tabular}
\end{table}

\subsection*{Probing the impact of spin and fidelity}
\begin{figure}[ht]
    \centering
    \includegraphics[width=\linewidth]{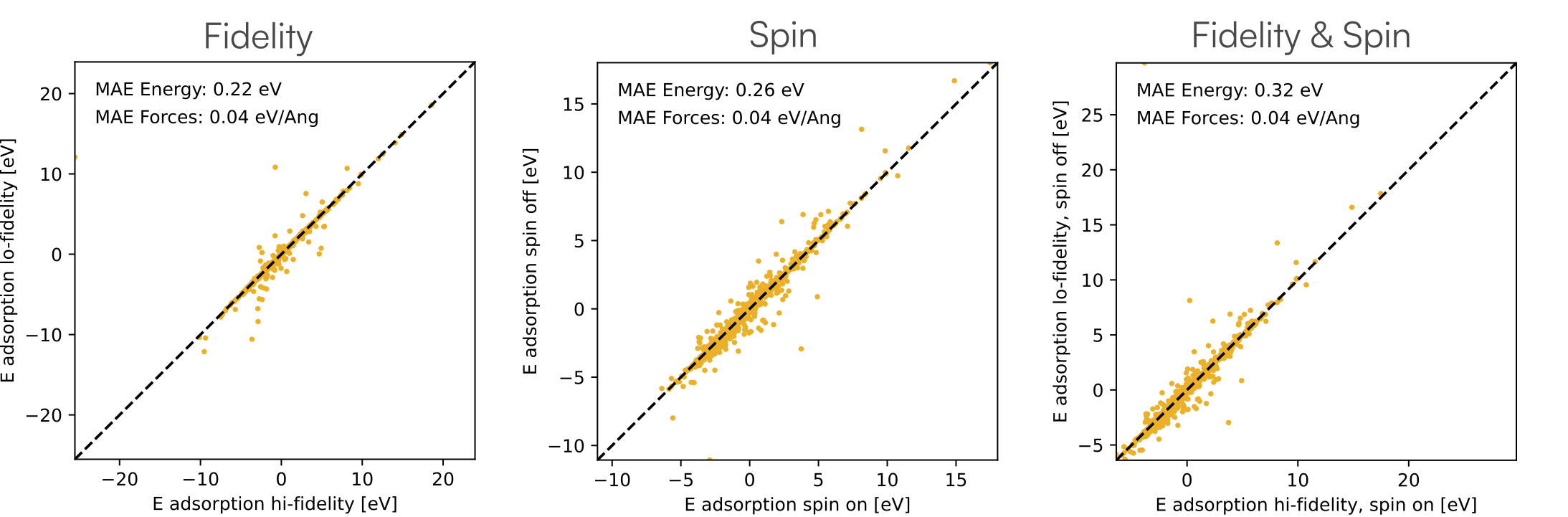}
    \caption{An investigation of the impact of fidelity being the same as OC20 (left), spin being off, rather than on (center), and both spin and fidelity being ablated simultaneously.}
    \label{fig:spin_fidelity_ablation}
\end{figure}

We also wanted to investigate the impact of spin and fidelity on the resultant energies. It is difficult to do this in a well posed way because the underlying bulk structure can be impacted by these DFT settings, so making a direct comparison is difficult. To attempt to do so, here we performed DFT single points on the DFT relaxed (with AQCat25 settings elections) adsorbate-slab configuration and the DFT relaxed slab (again with AQCat25 settings elections). The energies presented here are the difference between these two energies to exploit a cancellation of error from any differences in the true lattice constant. The single points were performed specifically ablating the settings highlighted. For fidelity (Fig. \ref{fig:spin_fidelity_ablation} - left), 500 spin-on systems and 500 spin-off systems were selected and single points were performed with ENCUT = 350 eV, and Methfessel-Paxton smearing with a width of 0.2 eV. For spin (Fig. \ref{fig:spin_fidelity_ablation} - center) 1000 systems with spin on were selected and single points were performed with spin off. For both spin and fidelity, 1000 systems with spin on were selected and single points were performed with the alternative fidelity and spin off. This can give some idea of the independent and combinatorial impact of these two factors on the DFT result.

\clearpage

\subsection*{Minimum adsorption energy task segmented by element category and spin category}
\begin{figure}[ht]
    \centering
    \includegraphics[width=\linewidth]{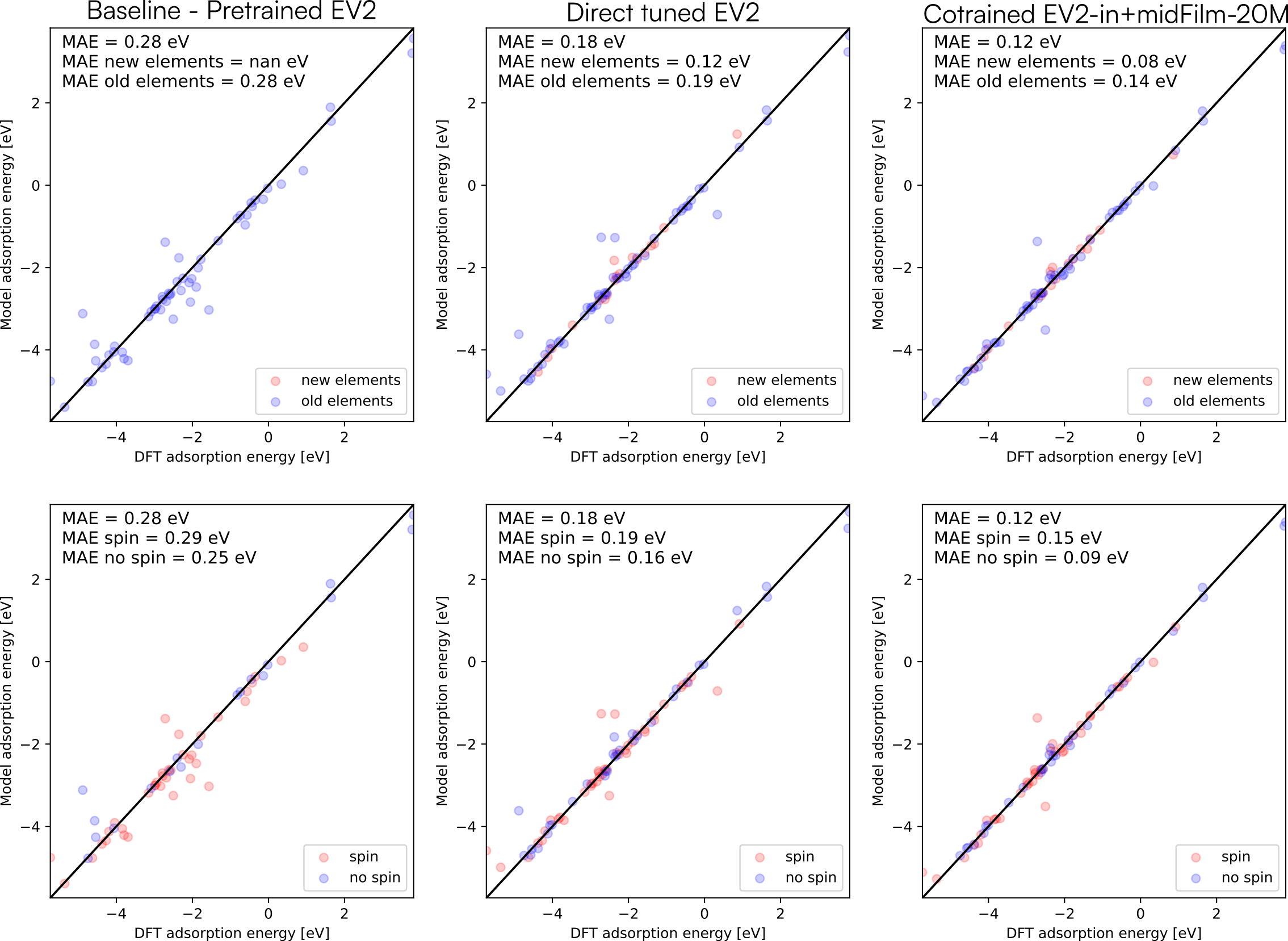}
    \caption{Model performance for finding the global minimum adsorption energy segmented by whether elements appear in OC20 (old elements) or not (new elements) - top and by whether the system was run with spin on or spin off - bottom.}
    \label{fig:adsorbml_supplementary}
\end{figure}

Figure \ref{fig:adsorbml_supplementary} splits the dense dataset results over whether the system was spin off or on and whether the system contains new elements, revealing that there are not any strong discrepancies between these groups. This is in alignment with Figure \ref{fig:material_magnetism}.

\clearpage

\subsection*{Material and spin splits when including organics}

\begin{figure}[ht]
    \centering
    \includegraphics[width=\linewidth]{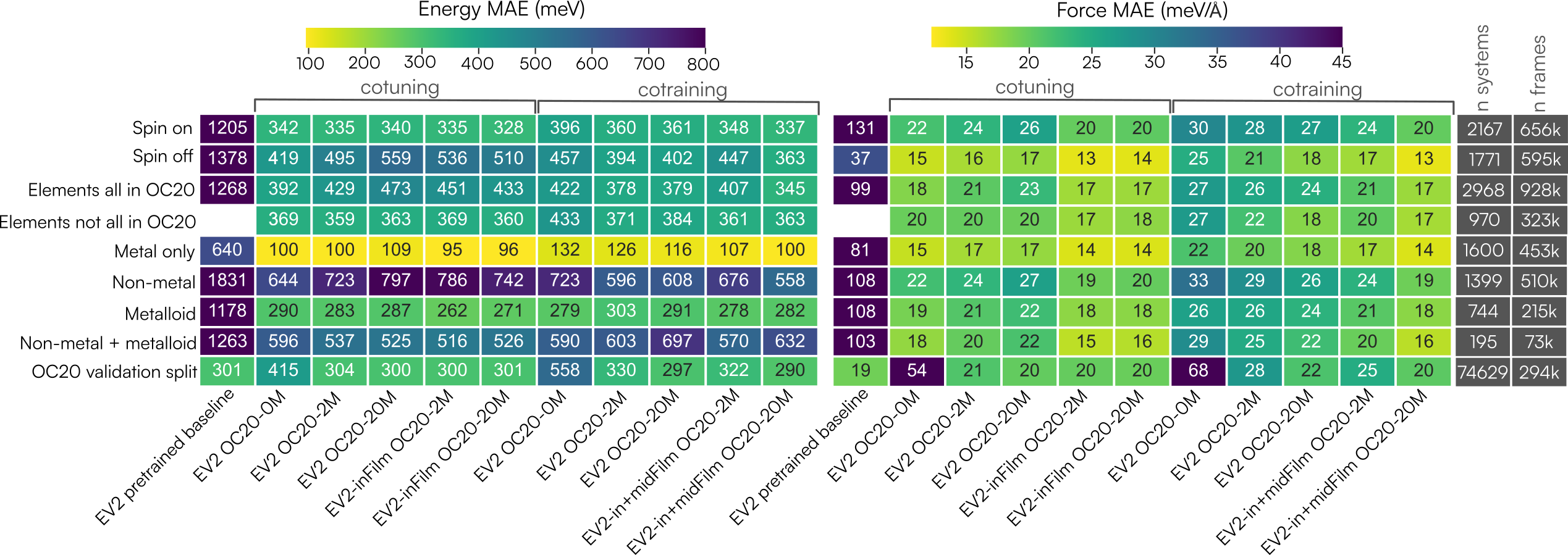}
    \caption{The same results presented in Figure \ref{fig:material_magnetism}, but without segregating the organic (fully non-metal) materials.}
    \label{fig:mat_mag_supp}
\end{figure}

When looking at the results including the organic materials, we see that the trends we would expect to see disappear. We would expect that performance on spin off systems should be better in most cases since that is the majority of data seen by the model, but because organic materials were all treated as spin off, the performance on spin off is dragged down. Metrics on non-metals are also pulled down. The same opposing trend is observed for new and in-OC20 elements. These organic materials will always be classed as in-OC20 element materials, and we see that performance is actually better on new elements because they drag down results for in-OC20 elements.

\clearpage

\subsection*{Additional looks at cotuning and cotraining energy metrics}

\begin{figure}[ht]
    \centering
    \includegraphics[width=\linewidth]{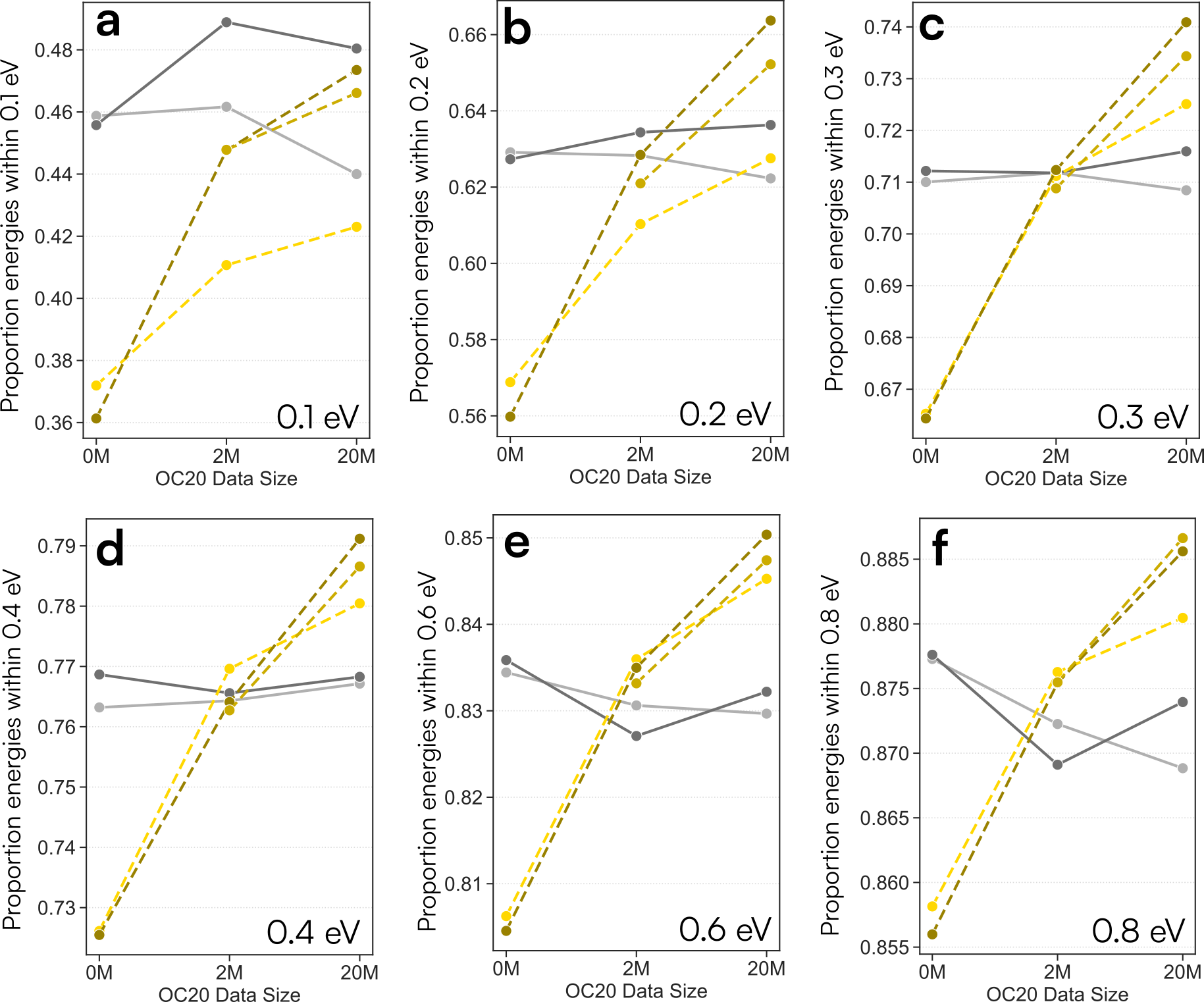}
    \caption{The evolution of trends with changing energy cutoff thresholds.}
    \label{fig:cotuning_metric_evo}
\end{figure}

\begin{figure}[ht]
    \centering
    \includegraphics[width=0.7\linewidth]{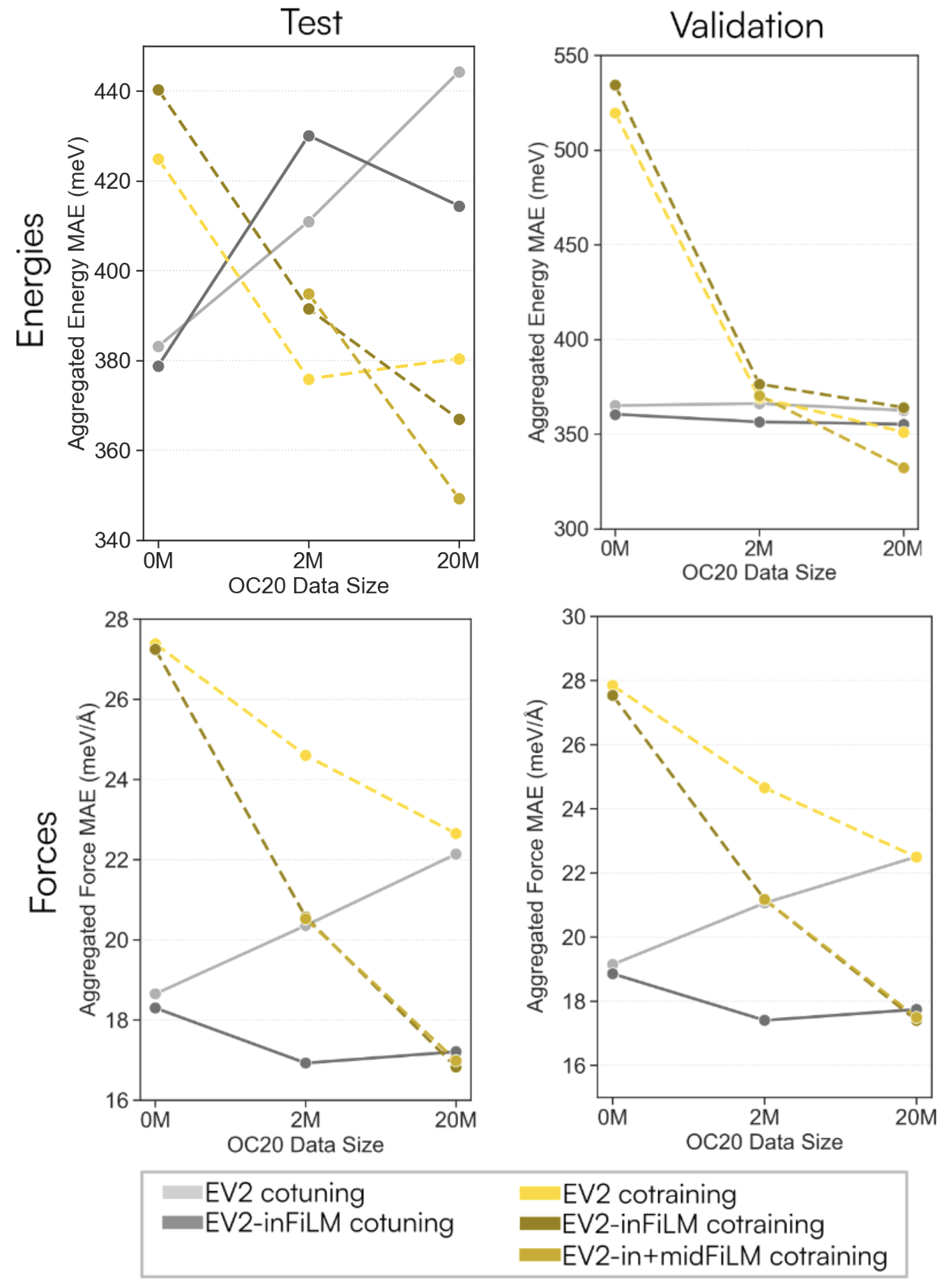}
    \caption{Performance of cotuned and cotrained models on test and val for energies (top) and forces (bottom).}
    \label{fig:cotunin_val_v_test}
\end{figure}

Figure \ref{fig:cotuning_metric_evo} shows the evolution of performance with changing energy cutoff. The cutoff is used to determine the proportion of systems with absolute energies errors less than the value. For the most strict cutoff, cotuning with FiLM has an advantage. For looser thresholds, cotraining has an advantage. Figure \ref{fig:cotunin_val_v_test} shows the energy and force MAE metrics on val and test for the cotuned and cotrained models. For forces, the trends are the same between the two. This is not true, however, for energies. This inspired us to investigate the cause which is that some very high energy errors are skewing the result. This is captured in Figure \ref{fig:material_magnetism}, which shows that performance is poor for organic materials. Cotraining models perform better on these materials at the expense of a slight reduction for other material classes. This shows that the trends for energy performance are sensitive to the metric selected. 

\end{document}